\documentstyle[preprint,aps,rotating,epsfig]{revtex}
\tighten
\newcommand{\tfrac}[2]{{\textstyle {#1\over #2}}}
\begin{document}
\bibliographystyle{revtex}
\draft
 \title{Nuclear Ground State Observables\\and QCD Scaling in a Refined\\
 Relativistic Point Coupling Model}
%\title{The Relativistic Point Coupling Model revisited}
\author{T. B\"{u}rvenich$^{1}$, D. G. Madland$^{2}$, J. A. Maruhn$^{1}$, and 
P. -G. Reinhard$^{3}$}
\address{$^{1}$Institut f\"ur Theoretische Physik, Universit\"at Frankfurt,
Robert-Mayer-Strasse 10,
D-60325 Frankfurt, Germany \\ $^{2}$Theoretical Division, Los Alamos National
 Laboratory,
Los Alamos, New Mexico 87544 \\ $^{3}$Institut f\"ur Theoretische Physik II,
 Universit\"at Erlangen-N\"urnberg, Staudtstrasse 7, D-91058 Erlangen, Germany}
\date{\today}

\maketitle

\begin{abstract}
We present results obtained in the calculation of nuclear ground state 
properties in relativistic Hartree approximation using a Lagrangian whose
QCD-scaled
coupling constants are all {\it natural} (dimensionless and of order 1).
Our model consists of four-, six-, and
eight-fermion point couplings (contact interactions) together with derivative
terms representing, respectively, two-, three-, and four-body forces and the
finite ranges of the corresponding mesonic interactions.
The coupling constants have been determined in a self-consistent procedure that
solves the model equations for representative nuclei simultaneously in a
generalized nonlinear least-squares adjustment algorithm.
The extracted coupling constants allow us to predict ground state properties
of a much larger set of even-even nuclei to good accuracy. The fact that
the extracted coupling constants
are all {\it natural} leads to the conclusion that QCD scaling and chiral
symmetry apply to finite nuclei.
\end{abstract}
\vspace{12pt}
 
\section{INTRODUCTION}
 
 Relativistic mean field (RMF) models are quite successful in describing
 ground state properties of finite nuclei and nuclear matter properties. 
They describe
 the nucleus as a system of Dirac nucleons that interact in a relativistic
 covariant manner via mean meson fields
 \cite{SW79,HS81,SW86,RRMGS88,R89,GRT90,FST97,LKR97,FR98}
 or via mean nucleon fields \cite{NHM92,HMMMNS94} whose explicit forms sometimes
 derive solely from the meson field approaches \cite{RF97}.
The meson fields are of finite range (FR) due to meson exchange whereas
the nucleon fields are of zero range (contact interactions or point
couplings PC) together with derivative terms that simulate
the finite range meson exchanges.
There are a number of attractive features in the RMF-FR and RMF-PC
approaches. These include
 the facts that the combined meson and/or nucleon fields account for the 
 effective
 central potentials that are used in Schr\"odinger approaches and that
 the physically correct spin-orbit potential occurs naturally with magnitudes 
 comparable to the
 (empirical) {\it ad hoc} spin-orbit interactions required in Schr\"odinger 
 approaches.
 Equally attractive is the fact that for relatively few parameters $(\simeq\: 
 10)$
 a vast amount of information is obtained:
 the Dirac single-particle wave functions and corresponding energy eigenvalues,
 the ground state mass, the baryon and charge densities together with their
 moments, and the properties of saturated nuclear matter.
 Furthermore, these quantities are obtained simultaneously
 in the same self-consistent relativistic Hartree (or Hartree-Fock) 
 calculation.\\
 
   In this work we use mean nucleon fields constructed with contact
 interactions (point couplings) to represent the system of interacting Dirac
 nucleons. We choose this approach for the following reasons: (a) possible
 physical constraints introduced by explicit use of the Klein-Gordon  
approximation
 to describe mean meson fields, in particular that of the (fictitious) sigma
 meson, are avoided and instead the effects of the various incompletely 
 understood and higher
order processes are assumed to be lumped into appropriate coupling constants
 and terms of the Lagrangian, as explained in Ref. \cite{NHM92}, (b)
 the use of point couplings  
 allows not only (standard) relativistic Hartree calculations to be performed,
 but also relativistic Hartree-Fock calculations \cite{MM88,MM89} by use of
 Fierz relations (up to fourth order \cite{MBM01}), and (c) the use of  
point couplings, because of their success in the Nambu-Jona-Lasino model
for the low-momentum domain of QCD \cite{KLE92},
is perhaps the best way to test for
 {\it naturalness} of the coupling constants in the seminal Weinberg expansion  \cite{We90}
 highlighting the role of power counting and chiral symmetry in weakening
 N-body forces. That is, two-nucleon forces are stronger than three-nucleon
 forces, which are stronger than four-nucleon forces, ..., resulting in a
 sequence making nuclear physics tractable. If the dimensionless
 coupling constants of the corresponding Lagrangian are of order 1 ({\it 
 natural})
 then QCD scaling and chiral symmetry apply to finite nuclei. Finally, (d), 
the RMF-PC model allows one to investigate its relationship to nonrelativistic
point-coupling approaches like the Skyrme-Hartree-Fock (SHF) approach
and the RMF-FR approach to contrast the importance and roles of the
different features these models have, as well as to obtain new insights. \\
 
Concerning point (c), the aim of this paper is to determine whether QCD scaling
and chiral symmetry apply to finite nuclei and, by their application,
to construct a state-of-the-art parameterization of the relativistic
mean-field point-coupling Lagrangian.
In the following we will use the term RMF model for both the version having 
finite range due to meson exchange, which we call RMF-FR, and the 
point-coupling (contact interaction) version that we denote by RMF-PC. \\
 
Concerning point (d), it is important to note here that one can also view RMF-PC as an approach
that lies in between the RMF-FR approach and the nonrelativistic
Skyrme-Hartree-Fock (SHF) approach which is also a well-developed self-consistent
mean-field model that performs very well (for a review see \cite{QUE78}).
Whereas SHF is based upon density-dependent contact interactions with
extensions to gradient terms, kinetic terms, and the spin-orbit interaction,
RMF-FR is based upon a coupled field theory of Dirac nucleons and
effective meson fields treated at the mean-field level, where density
dependence is modeled by nonlinear meson self couplings and the role
of gradient terms is taken over by the finite ranges of the mesons.
The kinetic and spin-orbit terms are automatically carried in both RMF
models \cite{R89}. Thus, a comparison of RMF-PC and SHF addresses the differences
between in-medium Dirac and Schr\"odinger nucleons, that is, in kinetic and spin-orbit
components, whereas a comparison of RMF-PC and RMF-FR addresses the
absence {\it vs.} presence of finite range and the different treatments
of density dependence.
Herein we will perform these comparisons using precisely the same fitting
strategy as in recent SHF and RMF-FR adjustments \cite{FRIREI,RRMGS88,REI95}
 except that here we will in addition be guided by
considerations of QCD scaling and chiral symmetry, that is, {\it naturalness}. \\

 We regard the present work with contact interactions as a refined
 relativistic point coupling model in comparison to our earlier work
 \cite{NHM92,HMMMNS94,FML96,DGM97} for the following three reasons.
 First, initial work in determining coupling constants in RMF-PC approaches
 \cite{NHM92,RF97} found a high correlation among the ground state observables
 used to determine them, particularly the total binding energy and the 
 root-mean-square
 charge radius. Given this fact and the presence of quadratic, cubic, and 
 quartic
 terms in the various densities appearing in the Lagrangian (representing two-,  three-,
 and four-body interactions) results in very delicate cancellations among the 
 corresponding
 many-body forces. This means that determination of the coupling constants
 using a nonlinear least-squares adjustment algorithm with respect to the 
 corresponding measured
 ground-state observables is fraught with difficulty because the coupling 
 constants are generally
 underdetermined. Consequently, the search for the minimum in the chi-squared
 hypersurface results in the location of many local minima from which erroneous
 conclusions can be drawn.
 Herein we address this problem more completely by applying two different 
 nonlinear least-squares
 adjustment algorithms and, finally, developing a combined adjustment algorithm
 that is used to determine our present results.
 Second, in our initial work we considered only spherical even-even closed-shell
 nuclei or closed-subshell nuclei in both proton number Z and neutron number N
 because, due to explicit omission of the pairing
 interaction, we allowed only orbital occupation probabilities of 0 or 1.
 Here, we introduce orbital occupation probabilities for both protons and 
 neutrons
 through a standard BCS approach in which the proton and neutron pairing
 strengths are simultaneously determined with the coupling constants in the
 adjustment algorithm.
 Third, most of our earlier work addressed the question of naturalness after the 
 fact,
 that is, without consideration of the complete set of 10 possible Lorentz 
 invariants
 that may occur (scalar, vector, pseudoscalar, axial vector, tensor, and the 
 same
 coupled to isospin $\vec\tau$) and without consideration of the QCD mass-scale
 ordering of the terms of the Lagrangian.
 The former consideration is neccessary to properly pursue the question of 
 naturalness
 while the latter consideration leads to, among other things, classification
 of the (allowed) terms of the Lagrangian according to their relative strengths,
 which, of course, assists in its construction in the first place.
 We address, and remain cognizant, of both of these considerations in the 
 various
 approaches presented here. \\

  The paper is structured as follows. The Lagrangian of our 
relativistic
point coupling model, together with its variants, is given in Sec. II.
Included are the corresponding relativistic Hartree equations, expressions
for the various densities and potentials appearing, and expressions for the
calculated observables that are to be used in determining the coupling 
constants of the Lagrangian.
The approximations that we invoke are also stated here.
In Sec. III we describe the determination of the coupling constants using four
different least-squares adjustment algorithms with respect to well-measured 
ground state observables and the external constraint of always obtaining 
reasonable calculated values
of the properties of saturated nuclear matter.
A relatively new approach to the $\chi^{2}$ minimization has been developed and
we explain how and why. Our results are given in Sec. IV. First, we present 
comparisons of calculation and experiment for nuclei whose measured observables 
were used to
determine the coupling constants. Second, we present comparisons of
predicted and measured
observables for nuclei not used in determining the coupling constants.
Third, we compare our results to those of other RMF approaches.
Then we give our final nuclear matter predictions and we mention initial 
results obtained in calculating fission potential energy surfaces and 
properties of
superheavy nuclei. We address the role of QCD scaling and chiral symmetry in 
Sec. V where we test our final sets of coupling constants for {\it naturalness} 
and present
the corresponding evidence obtained that QCD and chiral symmetry apply to 
finite nuclei.
Our conclusions and intentions for future work are given in Sec VI.
 
\section{THE MODEL}

\subsection{The Lagrangian}

The elementary building blocks of the point-coupling vertices are
two-fermion terms of the general type
\begin{equation}
  ( {\bar{\psi}} {\cal O}_\tau \Gamma  {\psi})
  \quad,\quad
  {\cal O}_\tau\in\{ {1},\tau_i\}
  \quad,\quad
  \Gamma\in\{1,\gamma_\mu,\gamma_5,\gamma_5\gamma_\mu,\sigma_{\mu\nu}\}
\end{equation}
with $\psi$ the nucleon field, $\tau_i$ the isospin matrices and
$\Gamma$ one of the 4x4 Dirac matrices.  There thus is a total of 10
such building blocks characterized by their transformation character
in isospin and in spacetime.\\

The interactions are then obtained as products of such
expressions to a given order.  The products are coupled, of course, to
a total isoscalar-scalar term.  By ``order'' we mean the number of
such terms in a product, so that a second-order term corresponds to a
four-fermion coupling, and so on. In second order only the ten
elementary currents squared and contracted to scalars may contribute,
but at higher orders there is a proliferation of terms because of the
various possible intermediate couplings.\\

In analogy to the nonrelativistic Skyrme-force models, one goes one
step beyond zero range and complements the point-coupling model by
derivative terms in the Lagrangian, as
e.g. $\partial_\nu\bar\psi\vec\tau_i\Gamma_j^\mu\psi$.  The
derivative is understood to act on both $\bar\psi$ and $\psi$.  The
derivative terms simulate to some extent the effect of finite range
and there may be genuine gradient terms from a density functional
mapping, as appears e.g. in electronic systems \cite{BLYP3}.\\
 
In the present work we consider the following four-fermion vertices:
 
\begin{center}
\begin{tabular}{lll}
 isoscalar-scalar:   &   $(\bar\psi\psi)^2$  
                     & \qquad ($\equiv$ $\sigma$-field)\\
 isoscalar-vector:   &   $(\bar\psi\gamma_\mu\psi)(\bar\psi\gamma^\mu\psi)$
                     & \qquad ($\equiv$ $\omega$-field)\\
 isovector-scalar:  &  $(\bar\psi\vec\tau\psi)\cdot(\bar\psi\vec\tau\psi)$
                    & \qquad ($\equiv$ $\delta$-field)\\
 isovector-vector:   &   $(\bar\psi\vec\tau\gamma_\mu\psi)
                         \cdot(\bar\psi\vec\tau\gamma^\mu\psi)$
                     & \qquad ($\equiv$ $\rho$-field)\\
\end{tabular}
\end{center}
\noindent
and their corresponding gradient couplings
$\partial_\nu(...)\partial^\nu(...)$.\\   
 
These constitute a complete set of second-order scalar and vector currents
whose coupling strengths in the corresponding Lagrangian we wish to
test for naturalness. All of them except for the derivative term for
isovector-scalar coupling have appeared in previous
RMF-PC work \cite{NHM92,HMMMNS94,FML96,DGM97,RF97}, however, the isovector-scalar
interaction ($\delta$-meson exchange) has been found not to improve the description
of nuclear ground-state observables in RMF-FR work \cite{R89,FST97}.
We therefore ask whether the insensitivity of the RMF-FR calculations
to the absence or presence of delta-meson exchange is due to
cancelations, other missing terms, unnaturalness, or a symmetry,
and we will investigate this same insensitivity in our RMF-PC
work here. That is, no RMF-PC calculation has yet included
simultaneously the four-fermion coupling plus corresponding
derivative for the isovector-scalar field, so we will do so
here. We postpone tensor couplings and 3rd and/or 4th order mixed
couplings [$(\bar{\psi}\psi)(\bar{\psi}\gamma_\mu\psi)(\bar{\psi}\gamma
^\mu\psi)$, for example] to our next work which will use the results
from this work as the starting point. For that work it is important  
to note that
whereas tensor couplings have had little effect in RMF-FR
calculations \cite{R89} they do have noticable effects in recent RMF-PC
calculations \cite{RF97}. Finally, the pseudoscalar channel ($\pi$-meson)
is not included here because it does not contribute at the
Hartree level.\\

In this work we begin with a set of higher-order terms that are
common to existing RMF-FR and RMF-PC studies. These are the familiar
nonlinear terms in the scalar coupling, $(\bar\psi\psi)^3$ and
$(\bar\psi\psi)^4$, as well as a nonlinear vector term
$[(\bar\psi\gamma_\mu\psi)(\bar\psi\gamma^\mu\psi)]^2$ as used in some
RMF-FR \cite{SUG94} and RMF-PC \cite{NHM92} models. Finally, of course,
the electromagnetic field and
the free Lagrangian of the nucleon field must be included.\\

Combining all of these terms, we obtain the Lagrangian of the
point-coupling model as

\begin{equation}
\begin{array}{lcl}
  {\cal L} 
  & = & 
  {\cal L}^{\rm free} + {\cal L}^{\rm 4f} + {\cal L}^{\rm hot}
  + {\cal L}^{\rm der} + {\cal L}^{\rm em},
\\[12pt]
  {\cal L}^{\rm free} \hfill
  & = & 
  \bar\psi ({\rm i}\gamma_\mu\partial^\mu -m)\psi,
\\[6pt]
  {\cal L}^{\rm 4f} \hfill
  & = & 
  - \tfrac{1}{2}\, \alpha_{\rm S} (\bar\psi\psi)(\bar\psi\psi)
  - \tfrac{1}{2}\, 
    \alpha_{\rm V}(\bar\psi\gamma_\mu\psi)(\bar\psi\gamma^\mu\psi)
\\[6pt]
  & &
   - \tfrac{1}{2}\, \alpha_{\rm TS} (\bar\psi\vec\tau\psi) \cdot
   (\bar\psi\vec\tau\psi)
  - \tfrac{1}{2}\,  \alpha_{\rm TV} (\bar\psi\vec\tau\gamma_\mu\psi)
    \cdot (\bar\psi\vec\tau\gamma^\mu\psi),
\\[6pt]
  {\cal L}^{\rm hot} 
  & = &  
  - \tfrac{1}{3}\, \beta_{\rm S} (\bar\psi\psi)^3 - \tfrac{1}{4}\, 
    \gamma_{\rm S} (\bar\psi\psi)^4 - \tfrac{1}{4}\, \gamma_{\rm V} 
    [(\bar\psi\gamma_\mu\psi)(\bar\psi\gamma^\mu\psi)]^2,
\\[6pt]
  {\cal L}^{\rm der} 
  & = & 
  - \tfrac{1}{2}\,\delta_{\rm S}(\partial_\nu\bar\psi\psi)
    (\partial^\nu\bar\psi\psi)  
  - \tfrac{1}{2}\,  \delta_{\rm V} (\partial_\nu\bar\psi\gamma_\mu\psi)
    (\partial^\nu\bar\psi\gamma^\mu\psi)
\\[6pt]
  & &
   - \tfrac{1}{2}\, \delta_{\rm TS} (\partial_\nu\bar\psi\vec\tau\psi) \cdot
   (\partial^\nu\bar\psi\vec\tau\psi)
  - \tfrac{1}{2}\, \delta_{\rm TV} (\partial_\nu\bar\psi\vec\tau\gamma_\mu\psi)
    \cdot (\partial^\nu\bar\psi\vec\tau\gamma^\mu\psi),
\\[6pt]
  {\cal L}^{\rm em} 
  & = & 
  -  e A_\mu\bar\psi[(1-\tau_3)/2]\gamma^\mu\psi -  
    \tfrac{1}{4}\, F_{\mu\nu} F^{\mu\nu}.
\end{array}
\label{eq:lagrang}
\end{equation}
\noindent
Note that we use the nuclear physics convention for the isospin where
the neutron is associated with $\tau_3=+1$ and the proton $\tau_3=-1$.\\

As it stands this Lagrangian contains the eleven coupling constants
$\alpha_{\rm S}$, $\alpha_{\rm V}$, $\alpha_{\rm TS}$, $\alpha_{\rm TV}$, $\beta_{\rm
S}$, $\gamma_{\rm S}$, $\gamma_{\rm V}$,
$\delta_{\rm S}$, $\delta_{\rm V}$, $\delta_{\rm TS}$, and $\delta_{\rm TV}$. The
subscripts indicate the symmetry of the coupling: ''S'' stands for
scalar, ''V'' for vector, and ''T'' for isovector, while the symbols
refer to the additional distinctions: $\alpha$ refers to
four-fermion terms, $\delta$ to derivative couplings, and $\beta$ and
$\gamma$ to third- and fourth order terms, respectively. \\

The model thus contains one or two free parameters more than analogous
RMF-FR models. This happens because most RMF-FR models make the tacit
assumption that the masses in the $\omega$- and $\rho$-field can be
frozen at the experimental values of the really existing mesons. The
assumption is justified to the extent that the actual fits to
observables are not overly sensitive to these masses.  In the RMF-PC model,
however, experience will still have to show whether the
derivative-term coefficients can be eliminated in a similar way, so
that for the present work all parameters are regarded as adjustable.

\subsection{The Mean-Field and No-Sea Approximations}

Similar to the RMF-FR approach, we consider the RMF-PC approach as an
effective Lagrangian for nuclear mean-field calculations at the  
  Hartree level without anti-nucleon states (no-sea
approximation). We thus obtain the mean-field approximation
\begin{equation}
   {\bar{\psi}}{\cal O}_\tau\Gamma {\psi}
  \longrightarrow
  \sum_{\varepsilon_\alpha>0}
   w_\alpha\bar\phi_\alpha{\cal O}_\tau\Gamma\phi_\alpha
\end{equation}
where the $w_\alpha$ are occupation weights to be determined by pairing,
see Sec. \ref{sec:pair}, the $\phi_{\alpha}$ are the Dirac four-spinor
single-particle wavefunctions with upper and lower components $g$ and $f$,
and the $\varepsilon_{\alpha}$ are the corresponding single-particle
energies. The ``no-sea''approximation is embodied
in the restriction of the summation to positive single-particle
energies. All interactions in the Lagrangian, Eq.
(\ref{eq:lagrang}), are then expressed in terms of the
corresponding local densities:
\begin{equation}
  \begin{array}{lrcl}
   \mbox{isoscalar-scalar:}
   & 
   \rho_{\rm S}(\vec{r})  
   &=&  \sum_{\alpha}\bar{\phi}_{\alpha}(\vec{r})\phi_{\alpha}(\vec{r}),
  \\[2pt]
   \mbox{isoscalar-vector:}
   & 
   \rho_{\rm V}(\vec{r}) 
   &=& 
   \sum_{\alpha} \bar{\phi}_{\alpha}(\vec{r})\gamma_0 \phi_{\alpha}(\vec{r}),
  \\[2pt]
   \mbox{isovector-scalar:}
   &
   \rho_{\rm TS}(\vec{r})
   &=&
   \sum_{\alpha}
   \bar{\phi}_{\alpha}(\vec{r})\tau_3\phi_{\alpha}(\vec{r}),
  \\[2pt]
   \mbox{isovector-vector:}
   & 
   \rho_{\rm TV}(\vec{r}) 
   &=& 
   \sum_{\alpha} 
   \bar{\phi}_{\alpha}(\vec{r})\tau_3\gamma_0\phi_{\alpha}(\vec{r}).
  \\[2pt]
   \mbox{proton:}
   & 
   \rho_{\rm C}(\vec{r})
   &=&
   \tfrac1{2}\left[\rho_{\rm V}(\vec{r})-\rho_{\rm TV}(\vec{r})\right].
 \end{array}
\label{eq:locdens}
\end{equation}

\subsection{Equations of Motion}

Following the mean field approximation, the single-particle
wavefunctions $\phi_\alpha$ remain as the relevant degrees of freedom.
Standard variational techniques \cite{BJODRE} with respect to $\bar\phi_\alpha$ yield
the coupled equations of motion for the set $\{\phi_\alpha\}$.
We are interested here in the stationary solution for which the
field equations read

\begin{equation}
\begin{array}{lcl}
  \gamma_0 \varepsilon_{\alpha}\phi_{\alpha}
  &=&
  ({\rm i} \vec\gamma\cdot\vec\partial  +  m + V_{\rm S} + V_{\rm V} \gamma_0
   + V_{\rm TS} \tau_3
   + V_{\rm TV} \tau_3 \gamma_0 
   + V_{\rm C} \tfrac{1-\tau_3}{2}\gamma_0) \phi_{\alpha},
\\[12pt]
  V_{\rm S} 
  & = & 
  \alpha_{\rm S} \rho_{\rm S} + \beta_{\rm S} \rho_{\rm S}^2+ 
  \gamma_{\rm S} \rho_{\rm S}^3 + \delta_{\rm S} \Delta \rho_{\rm S},
\\[6pt]
  V_{\rm V} 
  & = & 
  \alpha_{\rm V} \rho_{\rm V} + \gamma_{\rm V} \rho_{\rm V}^3
   + \delta_{\rm V} \Delta \rho_{\rm V},
\\[6pt]
  V_{\rm TS}
  &=&
  \alpha_{\rm TS}\rho_{\rm TS} + \delta_{\rm TS}\Delta\rho_{\rm TS},
\\[6pt]
  V_{\rm TV} 
  &=& 
  \alpha_{\rm TV}\rho_{\rm TV} +\delta_{\rm TV}\Delta\rho_{\rm TV},
\\[6pt]
  V_{\rm C}
  &=&
  eA_0,\quad \Delta A_0=-4\pi \rho_{\rm C},
\end{array}
\end{equation}

with the local densities as given in Eq. (\ref{eq:locdens}).

\subsection{Relation to Walecka-type RMF models}
\label{sec:preview}

The coupling between two nucleon densities is mediated by a finite
range propagator in the RMF-FR approach. We can expand the propagator
into a zero-range coupling plus gradient corrections in a standard
manner. This gives, for example, in the equation of motion for the
isoscalar-scalar coupling (see Eq. 2.10b of Ref. \cite{R89})
 
\begin{eqnarray}
\frac{-g_{\sigma}^{2}}{-\Delta + m_{\sigma}^{2}}\rho_{\rm S}
&\approx&
\underbrace{\frac{-g_{\sigma}^{2}}{m_{\sigma}^{2}}}_{\alpha_{\rm S}}\rho_{\rm S}
+
\underbrace{\frac{-g_{\sigma}^{2}}{m_{\sigma}^{4}}}_{\delta_{\rm S}}\Delta\rho_{\rm S}
\quad.
\end{eqnarray}
 
\noindent The identification with the corresponding parameters of the RMF-PC
is obvious by inspection of Eqs. (10-11). The reverse relations read
 
\begin{equation}
   m_{\sigma}^2
   =
   \alpha_{\rm S}/\delta_{\rm S}
   \quad,\quad
   g_{\sigma}^2
   =
   -\alpha_{\rm S}^2/\delta_{\rm S}
   \quad.
\end{equation}
 
\noindent Let us have a preview of what parameters to expect. Taking, for example, the set
NL-Z2 \cite{NLZ2} from the RMF-FR we
have $g_\sigma=10.1369$ and $m_\sigma=493.150~{\rm MeV}$. This yields an 
expected
$\alpha_S = -4.2252 \cdot 10^{-4}~{\rm MeV}^{-2}$ and $\delta_S = -1.7374 \cdot 
10^{-9}~{\rm MeV}^{-4}$. We will see in Section
\ref{sec:fitg} how the optimized RMF-PC parameterizations compare with
that. We shall also consider the similar relations belonging to the 
$\omega$-, $\delta$-, and $\rho$-meson, which are
 
\begin{eqnarray}
\alpha_{\rm V} & = & \frac{g_\omega^2}{m_\omega^2}\quad,\quad \delta_{\rm V} = 
\frac{g_\omega^2}{m_\omega^4} \\
\alpha_{\rm TS} & = & \frac{g_{\delta}^{2}}{m_{\delta}^{2}}\quad,\quad
\delta_{\rm TS} = \frac{g_{\delta}^{2}}{m_{\delta}^{4}} \\
\alpha_{\rm TV} & = &  \frac{g_\rho^2}{m_\rho^2} \quad,\quad \delta_{\rm TV} = 
\frac{g_\rho^2}{m_\rho^4} \quad.
\end{eqnarray}

\subsection{Pairing and the Center-of-Mass Correction}
\label{sec:pair}

In order to describe deformed and non-closed-shell nuclei reasonably,
pairing has to be involved. Since in this paper nuclei close to the
drip lines will not be considered, the pairing model can be kept
simple.  For reasons of better comparision, we use precisely the same
pairing recipe as in former RMF-FR calculations \cite{BEN00}. Thus we
employ BCS pairing with a $\delta$-force and use a smooth cutoff given
by a Fermi function in the single-particle energies. The occupation
amplitudes $u_\alpha$, $v_\alpha$ are determined by the gap
equation
\begin{equation}
 v_{\alpha}^2 = \frac{1}{2} \Bigg( 1 - \frac{\epsilon_{\alpha} - 
\epsilon_{\rm F}}{\sqrt{(\epsilon_{\alpha} - \epsilon_{\rm F})^2 + 
\Delta_{\alpha}^2}}\Bigg),~~~~~~v_\alpha^2 + u_\alpha^2 = 1,
\end{equation}
where the Fermi energy $\epsilon_F$ is to be adjusted such that the
correct particle number is obtained. The single-particle gaps $\Delta_\alpha$ 
are state dependent and are determined as
\begin{equation}
\Delta_\alpha = \int d^3 x\: \phi_\alpha^*(\vec{x}) \Delta(\vec{x})
\phi_\alpha(\vec{x}),
\end{equation}
where $\Delta(\vec{x})$ is the pair potential.
  The occupation weights for the
densities (\ref{eq:locdens}) are then given by $w_\alpha=v_\alpha^2$.
The pairing prescription introduces the pairing strength parameters
$V_{\rm P}$ and $V_{\rm N}$ for protons and neutrons, respectively. These are
fitted to experimental data simultaneously with the coupling constants
appearing in the Lagrangian. 

A further important ingredient is the center-of-mass correction and it
was shown \cite{BEN00ECM} that the actual recipe for that correction
has an influence not only on the predictions for light nuclei, but
also on the predictions for exotic nuclei.
Thus we use here 
the microscopic estimate
\begin{equation}\label{ecm}
E_{\rm cm}=\frac{ \langle \hat P^2 \rangle }{2 M},
\end{equation}
where $\hat P$ is the total momentum and $M$ the total mass of the
nucleus.  Again, we take care that the same recipe is used in the
RMF-FR forces to which we compare.

\subsection{Observables}
\label{sec:obs}

The determination of the coupling constants included in the model is
based upon a fit to experimental data, which requires precise numerical
comparison of calculated and experimentally observed quantities. For
mean-field models, the most natural quantities are bulk observables
such as the binding energy and the moments of the various density distributions,
while other properties such as the single-particle spectrum cannot be
related to experiment in a similarly precise way. In the following we
discuss the calculation of the observables used in our present study.

\subsubsection{Binding Energy}

The binding energy for a nucleus with $Z$ protons and $N$ neutrons 
is computed according to
\begin{equation}
  E_{\rm B}
  =
  Z m_{\rm p} +N m_{\rm n} -E_{\rm L}-E_{\rm pair}-E_{\rm cm}
  \quad.
\end {equation}
The most crucial part is the energy $E_{\rm L}$ from the mean-field 
Lagrangian. It is computed in standard manner as
$E_{\rm L}=\sum_\alpha v_\alpha^2 \epsilon_\alpha - \int {\rm d}^3x
\langle\Phi|{\cal L}|\Phi\rangle$
with $|\Phi\rangle$ being the BCS ground state. 
It can be rewritten in terms of the local densities as 
\begin{eqnarray}
   E_{\rm L}
   &=& 
   \sum_\alpha v_\alpha^2 \epsilon_\alpha
\nonumber\\ 
   &&
   - \int d^3x
   \biggl[ \tfrac1{2} \alpha_{\rm S}\rho_{\rm S}^2
   +\tfrac1{2} \alpha_{\rm V}\rho_{\rm V}^2
   +\tfrac1{2} \alpha_{\rm TS}\rho_{\rm TS}^2
   +\tfrac1{2} \alpha_{\rm TV}\rho_{\rm TV}^2
\nonumber\\
   &&
   \qquad
   +\tfrac2{3}\beta_{\rm S}\rho_{\rm S}^3
   +\tfrac3{4}\gamma_{\rm S}\rho_{\rm S}^4
   +\tfrac3{4}\gamma_{\rm V}\rho_{\rm V}^4
\nonumber\\
   &&
   \qquad
   +\tfrac1{2}\delta_{\rm S}\rho_{\rm S}\Delta\rho_{\rm S}
   +\tfrac1{2}\delta_{\rm V}\rho_{\rm V}\Delta\rho_{\rm V}
   +\tfrac1{2}\delta_{\rm TS}\rho_{\rm TS}\Delta\rho_{\rm TS}
   +\tfrac1{2}\delta_{\rm TV}\rho_{\rm TV}\Delta\rho_{\rm TV}
\nonumber\\
   &&
   \qquad
   +\tfrac1{2}\rho_{\rm C}V_{\rm C}
   \biggr]
   \quad.
\end{eqnarray}

\subsubsection{Charge Radius}

The starting point for all observables of the charge distribution is
the charge form factor defined by
\begin{equation}
  F_{\rm ch}(q) 
  = 
  \int d^3x~{\rm exp}\:(iq\cdot{\bf x})\rho_{\rm ch}({\bf x})
  \quad.
\end{equation}
It is a function of the momentum transfer $q=|{\bf q}|$ for spherically symmetric
charge distributions.  Note that the charge density is to be obtained
from folding the proton and neutron densities with the intrinsic
charge (and current) distributions of protons and neutrons. We use here the
same recipes as previously \cite{FRIREI,R89}. The various bulk properties
of the charge distributions are deduced in the standard manner \cite{FRI82}.
We calculate the r.m.s. charge radius then as
\begin{equation}
   r_{\rm rms}^{\rm ch} 
   = 
   \sqrt{-\frac{3}{F_{\rm ch}(0)} {\rm lim_{q\rightarrow 0}}
   \frac{d^2 F_{\rm ch}(q)}{d q^2}}.
\end{equation}

\subsubsection{Diffraction Radius}

The diffraction radius is obtained by comparison with a square-well
distribution having total charge Q:
\begin{equation}
\rho_{\rm box}^{\rm R} = Q\Big
(\frac{4\pi}{3} R^3\Big)^{-1},~~r\le R,~~~~ 0,~~r>R.
\end{equation}
Its formfactor is given by
\begin{equation}
F_{\rm box}^{\rm R}(q) = 3Q~\frac{j_1(qR)}{qR}
\end{equation} 
($j_1$ is the spherical Bessel function).  The diffraction radius $R$
is determined such that the first root $q_0$ of the actual formfactor coincides
with the first root of the box form factor.  This is the case
when $q_0 R$ is equal to the first root $x_0$ of $j_1(x)$.  We
obtain thus the diffraction radius by
\begin{equation}
R_{\rm dms} = \frac{x_0}{q_0} = \frac{4.4934095}{q_0}.
\end{equation}

\subsubsection{Surface Thickness}

The height of the first extremum for $q>0$ of the charge formfactor
can be used to obtain information about the characteristic surface
thickness $\sigma$ of the charge distribution. It
can be calculated by comparing the charge formfactor to a
distribution that is obtained by folding a box-like distribution
$\rho_{\rm box}^{\rm R}(r)$ over a Gaussian $\rho_{\rm g}^{\sigma}(r)$ with
width $\sigma$.  The formfactor $F_{\rm bg}(q)$ of the resulting
density distribution is simply the product of the single formfactors:
\begin{equation}
F_{\rm bg}^{\rm R,\sigma}(q) = F_{\rm box}^{\rm R}(q)F_{\rm g}^{\sigma}(q) = 
F_{\rm box}^{\rm R}(q)\:{\rm exp}\:\Big(-\frac{1}{2}\sigma^2 q^2\Big).
\end{equation}
The parameter $\sigma$ is determined in such a way that the original
charge formfactor and $F_{\rm bg}^{\rm R,\sigma}(q)$ have the same value
at the first extremum $q_e$ of $F_{\rm box}^{\rm R}(q)$. For $R$, the
diffraction radius as obtained in the previous subsection is used.
The final result for the surface thickness reads
\begin{equation}
\sigma = \frac{1}{q_e}\sqrt{2\:{\rm ln}\:\Bigg(\frac{F_{\rm b}^{\rm R_{\rm dms}}(q_e)}
{F_{\rm ch}(q_e)}\Bigg)}
\end{equation}
($q_e$ is given by $x_e/R_{\rm dms}$, $x_e$ being the x-value at the first 
extremum of $j_1(x)/x$).

\subsection{Comments on the Numerical Solution}

The coupled mean-field equations of the RMF-PC and RMF-FR models are solved on 
a grid in coordinate space employing derivatives as matrix multiplications in 
fourier-space. The solution that minimizes the energy of the system is 
obtained with the {\em damped gradient iteration} method \cite{DAMPEDGRAD} 
that has been successfully applied in the solution of such problems.

\section{DETERMINATION OF COUPLING CONSTANTS}
 
\subsection{The Task}

The RMF-PC model as presented above contains 11 coupling constants that
have to be determined in a fit to experimental data. We do this by a
{\em least-squares fit} i.e. minimization of
\begin{equation}
\chi^2 =\sum_i~\frac{(O_i^{\rm th}-O_i^{\rm expt})^2}{(\Delta O_i)^2},
\end{equation}
where $O_i^{\rm expt}$ are the experimental data and $O_i^{\rm th}$ denote the
calculated values. $\Delta O_i$ are the assumed errors of the
observables which empirically express the demands on the accuracy of
the model for the respective observables. They are in some cases (for
example for binding energies) larger than the experimental errors.
The observables chosen for the adjustment of the coupling constants
are discussed below, see section \ref{sec:obs}.  The global minimum of
$\chi^2$ should correspond to the optimal set of coupling
constants. Finding the global minimum is, however, a nontrivial and
non-straightforward task. To have a good chance to find the optimal
set of coupling constants, two different fitting algorithms have been
combined.

\subsection{The Fitting Algorithms}

The fitting of the model parameters is done with a combination of a
stochastic method and a direct method. The direct method is the {\em
Bevington Curved Step} \cite{BEV}, which had been successfully applied
in many previous fits within the RMF model \cite{RUF} and in the
Skyrme-Hartree-Fock model \cite{FRIREI,REI95}. It consists of the {\em
Levenberg-Marquardt} method \cite{NUMRE} with an additional trial step
in parameter space. The neighbourhoud of the supposed minimum is
modelled by a parabolic expansion. Close to the minimum this expansion
method is used, while further away the method switches to sliding down
along the gradient in parameter space. This method proved to be the
most effective one for the given problem and has been used for the
final minimization.\\

This fitting algorithm searches in each step for a lower
value of $\chi^2$. Thus it will walk towards the minimum which is
closest to the initial configuration in parameter space. This often
lets it get stuck in a local minimum. One way out is to start the
algorithm several times with different initial configurations. In
this work, a different way out has been chosen using {\em Simulated
Annealing {\rm (SA)}} \cite{NUMRE}.
This Monte Carlo algorithm does random steps in parameter space. For
each step, depending on an externally controlled temperature, the new
configuration is accepted or refused. At high temperatures, steps that
increase $\chi^2$ have a nonzero probability to be accepted. The more
the system is cooled down (we used exponential cooling), the smaller
the probabilities get for acceptance of a configuration which would
increase $\chi^2$. If the cooling is tuned to be sufficiently slow,
the routine can escape shallow minima and settle down in the global
minimum. Additionally, inspired by the techniques of {\em Diffusion
Monte Carlo} \cite{THI99}, a population of solution vectors (walkers)
has been used. Each walker, if unsuccessful in terms of $\chi^2$, has
a finite probability to die, while on the other hand successful
walkers can give birth to additional walkers that start their random
walk from that parent configuration.\\

Fig. \ref{mcrun} demonstrates a sample run (the set of observables for
this run is set 1 shown below). The maximum number of walkers was set
to 4. The two upper figures show the population walking through
parameter space in the directions $\alpha_{\rm S}$ (left) and
$\alpha_{\rm V}$ (right).  The strong correlations among these two
parameters are nicely illustrated in correlated movements of the
walkers.  The figures below show the $\chi^2$ for the different
configurations as well as the external temperature and the number of
walkers.  Due to failures of some walkers, the population decreases
(walkers die after around 1600 iterations ). This information is fed
back into the algorithm, so that at later times the chance for new
walkers to be born increases again: after  additonal 900 iterations
new walkers are born.

Using that population of walkers (two runs with a maximum number of 4
walkers were performed) increases the chance to find the desired
minimum of $\chi^2$. Simulated annealing is, however, quite slow
compared to the other routine and needs considerably more iterations
than the direct methods.\\

To get the best from both classes of fitting algorithms, the following
mixed fitting procedure was used to determine the coupling constants:
Two populations of parameter vectors were propagated through parameter
space with simulated annealing, using the smaller set of observables
denoted below as set 1.  After the system had been cooled down
sufficiently, these parameter vectors were used as starting points for
the {\em Bevington Curved Step} procedure using the larger set 2 of
observables.

\subsection{The Choice of Observables}
\label{sec:obs}

Two different sets of observables were employed due to the different
aim of the minimization procedures.  To explore the parameter space
for minima with SA, set 1 was used, which is identical to the
set of observables used for the determination of the coupling
constants in Ref.\cite{NHM92} (see table \ref{nuc-set-1} for the
observables and weights).  Since the idea here was to locate basins
around minima, that set of observables proved to be sufficient to
indicate these areas.\\

For the final minimization, however, the larger set 2 (see table
\ref{nuc-set-2}) was used.  The pairing strenghts were adjusted
simultanously, which is important in order to obtain a set of coupling
constants with predictive power comparable to other mean-field
approaches.  This set of observables had successfully been applied
before to fit the parameterization NL-Z2 for the RMF model \cite{NLZ2}
(the only exception being that the pairing strenghts for NL-Z2 used for
the calculations in the present work
were
adjusted with a larger set of empirical gaps, see Ref. \cite{BEN}).

\subsection{The Force PC--F1}
\label{sec:fit1}

Our first RMF-PC calculation is that corresponding to RMF-FR
approaches which treat exchange of the $\sigma$, $\omega$, and $\rho$
mesons. Thus we have three linear terms together with three corresponding
derivative terms and three higher order terms.
The set of nine coupling constants emerging from the fitting procedure with
the lowest value of $\chi^2$ is called PC--F1 and is shown in Table
\ref{p-f1}.

Note that these coupling constants have correlated errors and
uncorrelated errors \cite{BEV}. The uncorrelated error of a parameter
is the allowed variation of that isolated parameter (while all other
parameters are kept fixed) which enhances $\chi^2$ just by the value
one.  The parameters have thus to be given with enough digits that the
last digit stays below the uncorrelated error. This rule is obeyed in
the above table. The correlated error of a parameter is its allowed
change, i.e. within $\chi^2+1$, if all the other parameters are
readjusted. Correlated and uncorrelated error would be the same if a
parameter is completely independent from the others.  In practice, the
correlated errors are much larger than the uncorrelated ones,
indicating strong correlations among the parameters. The largest
correlations appear between $\alpha_{\rm S}$ and $\alpha_{\rm V}$
whose sum happens to provide the largest contribution to the nuclear
shell model potential. 
The correlated error of $\delta_{\rm TV}$ is quite large and shows
that the parameter might as well have a positive or zero value. It is
quite losely determined by the fitting strategy. There is an analogous
situation for Skryme forces, where some of the isovector terms posses
only losely determined parameters.
The pairing strengths, on the other hand, have small discrepancy
between correlated and uncorrelated error. This shows that the fit to
pairing is basically independent from the fit of the mean field
properties.

\subsection{Quality of the Fit}

The total $\chi^2$ for PC--F1 is $99.1$. Additionally, we consider the 
$\chi^2$ per point, $\chi^2_{\rm pt} = 2.11$, where the number of points is 
the number of observables taking into account in the fitting procedure, 
which is 47 in our case. The $\chi^2$ per degree of freedom, 
$\chi^2_{\rm df} = 2.75$, where the degrees of freedom are caluclated as 
the difference between data points and the number of free parameters, 
also measures the quality of the force obtained in the fitting procedure.
These number need a bit more
eludication. To that end, we inspect the (dis-)agreement for the
various fit observables in detail. This is done in
Fig. \ref{fitnuclei} which demonstrates performance of the new RMF-PC
force PC--F1 and compares it to the RMF-FR force NL-Z2.
One sees that the binding energy is described most
precisely with an average accuracy of $0.25\%$.  The radii are
reproduced within about $0.5\%$ percent. The surface thickness comes
last. But mind that the usage of relative errors punishes this
quantity which has a comparatively low value of $1\,{\rm fm}$.
Most actual errors stay within these error bands. There are a few
exceptions. The energies of $^{40}$Ca and Ni isotopes seem to 
have trouble and the diffraction radius of $^{112}$Sn is a bit large. 
Comparing the average errors between PC--F1 and NL-Z2, we see slightly
different trends.  NL-Z2 is superior with respect to binding energies
and surface thicknesses.  It does, however, perform less well
concerning radii.  The total $\chi^2$ of NL-Z2 is $132.7$ which is $34\%$ 
larger than that for PC--F1. The overall
performance of the point-coupling thus seems to be a bit better, although
the difference is not too dramatic. 

\subsection{Exploring Modifications in the Isovector Channel}

The model Lagrangian (\ref{eq:lagrang}) contains only a
bare minimum of isovector terms. This was chosen in close analogy to
the RMF-FR. There are many more terms conceivable already at the given
order of couplings. The problem is that the given obervables all
gather around the valley of stability and contain only little
isovector information. Isovector extensions of the model are thus not
so well fixed by the data. Nonetheless, it is worth exploring those
extensions in order to check that one is not missing too much in the
above standard model.

\subsubsection{Isovector-Scalar Terms}

We now test the linear isovector-scalar term with coupling constant
$\alpha_{\rm TS}$ [see Eqs. (4) and (13)].
Table \ref{alphats} shows the set of ten optimized coupling constants
which we call PC--F2.
The correlated errors of the isovector coupling constants
are much larger than in PC--F1 (see Table \ref{p-f1}). The $\chi^2$ for
the extended set given was reduced by only $0.6\%$ compared to
PC--F1. Thus we find that this extension is not well determined by
the present set of data.  It is interesting to note that the sum of
$\alpha_{\rm TS}+\alpha_{\rm TV}$ approximately corresponds to the
value of $\alpha_{\rm TV}$ in the force PC--F1. This may indicate that the
overall isovector strength has a well defined value, but the detailed
splitting between the two terms is not yet well determined.

\subsubsection{Nonlinearities in the Isovector-Vector Terms}

Another obvious extension of the model is the lowest order nonlinear term in the
isovector-vector density,
\begin{equation}
  {\cal{L}}^{\rm hot}_{\rm TV} 
  = 
  -\frac{1}{4} \gamma_{\rm TV} 
  [(\bar{\psi}\vec{\tau}\gamma_{\mu}\psi) 
   \cdot (\bar{\psi}\vec{\tau}\gamma^{\mu}\psi)]^2
\quad.
\label{eq:pcf2}
\end{equation}
The ten optimized coupling constants which we call PC--F3 are shown in Table 
\ref{gammatv}.
The new parameter $\gamma_{\rm TV}$ is characterized by large
uncorrelated and correlated errors, and in addition the uncertainties
in the parameter $\delta_{\rm TV}$ have increased compared to the
force PC--F1. This hints that the experimental observables are unable to pin down
the new parameter. The overall quality is $\chi^2=98.8$ which is only
$0.3\%$ better than that of PC--F1. This indicates that 
the extension by a nonlinear isovector term is undetermined at the
present stage of the fits.

\subsubsection{An extended set with 11 coupling constants}

As a last test of possible extensions in the isovector-channel we performed a 
fit including the four isovector parameters $\alpha_{\rm TS}, \delta_{\rm TS},
\alpha_{\rm TV}, \delta_{\rm TV}$. 
The emerging set of eleven coupling constants is shown in Table \ref{p-f4} and 
is called PC--F4. This set has a small negative coupling constant in front of 
the four-fermion isovector-scalar term leading to a small attraction. 
The sum $\alpha_{\rm TS} + \alpha_{\rm TV}$ leads to a value of $\approx 3.3 \times 10^{-5}~{\rm MeV}^{-2}$, 
which is quite close to the value obtained for $\alpha_{\rm TV}$ in the 
set PC--F1. This observation underlines the statement we have already made 
concerning the force PC--F2, where we saw a similar behavior of the extended 
isovector strength. Due to the large correlated errors, all isovector parameters 
except $\alpha_{\rm TV}$ are compatible with positive or zero values, showing 
that the isovector channel of this effective Lagrangian is not well determined 
by the data included in the fit.

\subsection{Comparison with Walecka-type models}
\label{sec:fitg}

In Section \ref{sec:preview}, we estimated expected coupling constants from a
gradient expansion of the finite ranges in RMF-FR.
We compare now the values for the various coupling constants with values that 
we can expect from the finite-range RMF model, choosing the interaction 
NL--Z2 for our comparisons.
Table \ref{reltowalecka} shows the expected values (except for the 
isovector-scalar channel, since the RMF model with NL--Z2 has no 
$\delta$ meson) together with the values taken from NL-Z2.

Good agreement can be seen for the coupling constants mainly responsible for the 
nuclear potential, namely, $\alpha_{\rm S}$ and $\alpha_{\rm V}$, which 
are very similar in each of the RMF-PC forces and are somewhat lower than 
the corresponding RMF-FR values.\\ 

By looking at the results for the corresponding coupling constants 
$\delta_{\rm  S}$ and $\delta_{\rm V}$, we realize that there are dramatic
discrepancies. In none of the interactions does the sign of $\delta_{\rm V}$ agree 
with the RMF-FR value. 
Only in PC--F4 do all signs of the four isovector coupling constants
comply with the expectations from the estimates [Eqs. (19) and (20)].
One has to keep in mind, however, that these coupling constants, due
to their large correlated errors, are not incompatible with zero.
The values for $\alpha_{\rm TV}$ agree well with the expected value from 
NL--Z2, reflecting about the same asymmetry energy that all RMF-FR forces 
deliver (see the discussion about nuclear matter properties in the next 
section). \\

One may be suspicious that the different mapping of non-linearities
spoils the comparison. To countercheck, we performed one more fit
including the 11 coupling constants, but setting $\gamma_{\rm V}$ to
zero in order to address the different signs of  
$\alpha_{\rm V}$ and $\delta_{\rm V}$ which appear in all sets of
coupling constants studied. The resulting set of coupling constants
still has the same signs, which shows that the
negative value of $\delta_{\rm V}$ is not related to having
nonlinearities in the isoscalar-vector channel of the effective
Lagrangian.
We thus are led to the conclusion, that the gradient terms in the
RMF-PC embody obviously more than just a compensation for the finite
range. This may indicate that the present RMF-PC Lagrangian is incomplete.
\\
 
Altogether, all isovector extensions turned out to improve the fits
only very little. Even a detailed analysis of the trends along
isotopic chains did not show any significant improvement. Thus,
we did not consider additional forces in our present study
because they do not appear to be well determined with existing observables.
Additionally, the $\chi^2$ per degree of freedom is larger for the 
extended sets compared to PC--F1, showing that at the present stage the 
extended forces do not incorporate real physical improvements.
This may change for larger sets of observables which include dedicated
isovector data. The large
uncertainties in the isovector coupling constants in the three extended models
shows that there is indeed sufficient freedom to accomodate new
isovector observables.

\section{RESULTS}
 
\subsection{Comparisons}

We now check the predictive power of the newly fitted
force PC--F1. We do this by looking at the performance for a variety of
test cases and observables which were not included in the fit.  We
compare the model both to experimental data and to three other
relativistic mean-field approaches, namely the older point-coupling
model PC-LA \cite{NHM92}, and the two sets NL3 \cite{LAL97} as well as
NL-Z2 \cite{NLZ2} from the family of RMF-FR models.  NL-Z2 had been
fitted with precisely the same set of data as PC--F1. PC-LA employed a
smaller set of data as discussed above.  NL3 was fitted to binding
energies, charge radii and neutron rms radii of the nuclei $^{16}$O,
$^{40,48}$Ca, $^{58}$Ni, $^{90}$Zr, $^{116,124,132}$Sn and
$^{208}$Pb. Additionally, nuclear matter properties entered into the
fit (E/A = $-16~$MeV, $\rho_0$ = $0.153~$fm$^{-3}$, $K$ = $250~$MeV,
$\alpha_{\rm sym}$ = $33~$MeV).  NL-Z2 and NL3 are two state-of-the-art
mean-field forces that have been tested in a variety of
applications. So this selection of forces will give us a well-balanced
picture of the quality of modern relativistic mean-field forces.  In
some cases we compare also with state-of-the-art Skryme forces, namely
the forces SLy6 \cite{CHA95} and SkI3 \cite{REI95}. SLy6 aims at
describing extremely neutron-rich systems up to neutron stars together
with normal nuclear matter and nuclei.  SkI3 has a spin-orbit force
that in its isovector properties is analogous to the nonrelativistic
limit of the RMF-FR model and was fitted using the strategy of
Ref. \cite{REI95} which is much similar to the strategy and input data
used here.

\subsection{Nuclear Matter}

Table \ref{nucmat} shows the bulk properties of symmetric nuclear
matter as predicted by the different forces.  Like the other RMF
approaches, PC--F1 has a rather low saturation density of around
$\rho_0 = 0.15 ~{\rm fm}^{-3}$ while the Skyrme forces produce the
larger $\rho_0 = 0.16 ~{\rm fm}^{-3}$ (which is close to the commonly
accepted value \cite{MYERS}). Additionally, all RMF forces favor a
larger binding energy at the saturation point. These are systematic
differences between the two approaches apparent for both types of RMF
as compared to SHF. This indicates that these trends are not due to a
finite-range in RMF-FR but must have other reasons related to
relativistic kinematics.\\

The incompressibility $K$ of PC--F1 is comparable to that of PC-LA and
NL3, whereas NL-Z2 produces a much smaller value.  The larger value of
$270\,{\rm MeV}$ is much closer to the commonly accepted $240\,{\rm
MeV}$ while the value of NL-Z2 is far too small.  It is interesting to
note that the large value of $K$ was aimed at in the fit of NL3 while
it just emerged from the fit for PC--F1. It is also to be remarked that
NL3 achieves this large $K$ at the price of producing somewhat too
small surface thickness. PC--F1, on the other hand, describes surface
thickness as well as NL-Z2 (see Figure \ref{fitnuclei}) and has a much
larger $K$ than NL-Z2. We see here a clear difference of the
point-coupling versus finite range. This is corroborated by the fact
that the SHF models are also point-coupling models and do also tend to
predict incompressibilities in the range of $250\,{\rm MeV}$.\\

The symmetry energy $a_{\rm sym}$ has the same large value in all
RMF models while SHF results stay closer to the commonly accepted
values ($\approx 30\,{\rm MeV}$). This is a systematic discrepancy
between RMF and SHF. It is most probably connected to the rather
rigid parameterization of the isovector channel in RMF.

The effective mass is consistently small in all RMF models while SHF
can cover a broad range of values up to $m^*/m=1$ and even a bit
larger, see e.g. \cite{TON00}. The value of $m^*/m$ in the RMF is directly 
related to
the strength of the vector and scalar fields which, in turn,
determines the spin-orbit splitting. There is thus little freedom to
tamper with the effective mass in RMF unless one alternative means to
tune the spin-orbit force (as e.g. a tensor coupling). \\

Fig. \ref{nucmatfig} shows several features of symmetric nuclear
matter as function of density $\rho$. The results are about similar
for NL-Z2, NL3, and PC--F1 while PC-LA shows dramatic deviations,
particularly for $\rho>0.17\,{\rm fm}^{-3}$. The net potential
$V=V_S-V_V$ and the effective mass $m^*$ play a crucial role to
determine the spectra of finite nuclei. Thus we have to expect
somewhat unusual spectral features for PC-LA. At second glance, we see
also slight differences between the other parameterizations coming up
slowly at larger densities. The equation of state $E/A$ is less rigid
for PC--F1 (correlated with a slightly smaller potential $V$ and less
suppressed $m^*$). This is consequence of the fact that the density
dependence is parameterized differently in point coupling and finite
range models.

\subsection{Neutron Matter}

Neutron matter is a most critical probe for the isovector
features. It has been exploited extensively in the adjustment of the
SHF forces \cite{CHA95}. There are, of course, no direct
measurements. But neutron matter is well accessible to microscopic
many-body theory such that there exist several reliable calculations
of its properties. 
Figure \ref{neutronmatter} shows the equation of state for the four
RMF forces and SLy6. The crosses correspond to data from
\cite{FRI81}.  
We confine the comparison to low densities which are relevant for
nuclear structure physics. It is obvious that all RMF models show a
similar trend which, however, differs significantly from the ``data''
and from SLy6. This is a systematic discrepancy which, again, is
related to the rather sparse parameterization in the isovector
channel.

\subsection{Binding Energies}

\subsubsection{Isotopic and Isotonic Chains}

In Figures \ref{isotopes} and \ref{isotones}, we show the systematics
of relative errors on binding energies along isotopic and isotonic
chains for the two RMF-PC forces and the RMF-FR forces discussed
here. All nuclei in these figures are computed as being spherical.
Note that the scales are different for each figure.  As guideline we
indicate by horizontal dotted lines the average error of the models for this 
observable. \\

Larger errors show up sometimes for small nuclei in the isotopic
chains, see Figure \ref{isotopes}. The case $^{40}$Ca is notoriously
difficult for PC--F1 and small Ni isotopes are a problem for all RMF
models.
The underbinding of $^{40}$Ca may be excused by a
missing Wigner energy \cite{SATWYS}.  But $^{56}$Ni is already
overbound and a Wigner energy would worsen the situation. The reasons
for the deviation have to be searched somewhere else, probably it is
again an isovector mismatch.\\

The heavier systems perform much better. They are described within an
error of about $0.4\%$, with few exceptions.  We also see that NL-Z2
performs best in most cases.  Some slopes and kinks are also apparent
in these plots for all forces. They indicate yet unresolved isotopic
and isotonic trends.  Another interesting observation can be made: the
structure of the curves is, with differences in detail, similar for
NL-Z2 and PC--F1 in almost all cases (this is most striking for the Sn
isotopes). It shows that the fitting strategy (i.e. the choice
of nuclei and observables) has direct consequences for the 
trends of the errors.\\

A well visible feature are the kinks of the errors which appear at
magic shell closures. These kinks indicate that the jump in separation
energies at the shell closure is too large (typically by about 1-2
MeV). This, in turn, means that the magic shell gap is generally a bit
to large. Some SHF forces solve that problem by using effective mass
$m^*/m=1$. This option does not exist in RMF as we have seen
above. But there are other mechanisms active around shell
closures. The strength and form of the pairing can have an influence
on the kink ($\equiv$ shell gap). Moreover, ground state correlations
will also act to reduce the shell gap of the mere mean field
description. This is an open point for future studies. \\
 
Figure  \ref{isotones} show the relative errors of binding along
isotonic chains, assuming again all spherical nuclei. 
Again, there are larger fluctuations for the small nuclei $N=20$ and
$N=28$ while the heavier $N=50$ and $N=82$ stay nicely within the
error bounds. But the heaviest $N=126$ chain grows again out of bounds
at its upper end. Isotonic chains are a sensitive test of the balance
between the Coulomb field and the isovector channel of the effective
Lagrangian. All effective forces discussed here produce larger errors
compared to the experimental isotonic chains, which shows the need
for further investigations of this property of the RMF models.

\subsubsection{Superheavy Elements}

The upper panel of Figure \ref{heaviestgg} shows the relative errors
in binding energies for the heaviest even-even nuclei with known
experimental masses (compare with a similar figure in
Ref. \cite{BUE98}). The lower panel delivers as complementing
information the ground-state deformations expressed in terms of the
dimensionless quadrupole momentum $\beta_2$.  The calculations were
performed with allowing axially symmetric deformation assuming
reflection-symmetric shapes. The agreement is remarkable. All forces
(with some exceptions for PC-LA) produce only small deviations which
stay well within the given error band. This is a gratifying surprise because 
we are here
40-50 mass units above the largest nucleus included in the fit. It is
to be noted that most SHF forces do not perform so well and have
general tendency to underbinding for superheavy nuclei \cite{BUE98}.
There are also (small but) systematic differences between the RMF
models. NL3 generally overbinds a little while NL-Z2 and PC--F1 tend to
underbind.  All forces show yet unresolved isovector trends. The
increase of the binding energy with increasing neutron number is too
small.  These trends were already apparent for known nuclei (see the
discussion above).  The reasons for all these trends are not yet
understood.  Finally, mind the kinks visible for the $Z=98$ and
$Z=100$ isotopes at neutron number $N=152$ which hint at a small
(deformed) shell closure there.\\

All forces predict strong prolate ground-state deformations for these
superheavy nuclei ($\beta_2 \approx 0.26-0.31$). The trends look
similar for all forces. The largest deformations appear at $N=148$
and/or $N=150$.  But there are systematic differences in detail: NL-Z2
has always larger ground-state deformations than the other
forces, while PC--F1,
PC-LA, and NL3 show much similar deformations. The difference is probably
related to the surface energy:  NL-Z2 has a lower surface energy than
NL3.  The symbol with error bars at Z/N=102/152 in Figure \ref{heaviestgg} corresponds
to the measured ground-state deformation of $^{254}{\rm No}$
\cite{REI99,LEI99}. This deformation is overestimated by all forces,
PC-LA and NL3 stay within the error bars, though.
The error ranges from 6 to $13\%$ which is still acceptable.

\subsection{Fission Barrier of $^{240}$Pu}

Fig. \ref{pu240} shows the fission barrier of $^{240}$Pu computed in
axial symmetry allowing for reflection asymmetric shapes (for a
discussion of the numerical methods, see reference \cite{RUT96}). The
experimental values for ground state deformation, barrier, and isomer
energy are taken from Ref. \cite{BJO80,FIR96,LOB70,BEM73}.  All forces
predict the same ground-state deformation in agreement with the
experimental value and they all show the typical double humped
structure of the fission barrier. Also the first barrier (which
corresponds to reflection-symmetric shapes) is very similar but too
large as compared to experiment.  That may be a defect of symmetry
restrictions.  Triaxial degrees of freedom can decrease the calculated
barrier by about 2~MeV \cite{RUT96}, which would bring the curves
closer the the experimental value.  Moreover, the (yet to be calculated)
zero-point energy corrections will also lower the barriers somewhat
  \cite{Rei79}.  \\

Larger differences develop towards the second minimum and further out
(where also the asymmetric shapes take over).  This can be related to
the surface properties of the different forces.  Forces with a high
surface energy place the isomeric state higher up than forces with
lower surface energy.  All forces, however, underestimate the
experimental value for the energy difference of the ground-state and
the isomeric state, which is 2.3~MeV. Vibrational zero-point energies
may still help in case of NL3. But the minima for the other three
forces are to deep that those small corrections could bridge the gap.

\subsection{Observables of the Density}

\subsubsection{Charge Radius, Diffraction Radius and Surface Thickness}

In this section we take a look at the observables which are related to
the nuclear charge distribution, the r.m.s. and diffraction radii as
well as the surface thickness (see Sec. \ref{sec:obs}).  In Figure
\ref{densrelated} we show results for Sn and Pb isotopes. The experimental
data are taken from Ref. \cite{FRI82,OTT89,FRI95}.\\

The r.m.s. and diffraction radii are described generally good.  PC-LA
yields to large diffraction radii in Sn isotopes. NL-Z2 produces a bit
to large radii in Pb isotopes. But note that all forces reproduce the
trends of the r.m.s. radii in lead with its pronounced kink at the
magic $N=126$. It is a known feature RMF-FR models perform very well
in that respect \cite{SHALAL,REI95} and we see here that the
point-coupling models maintain this desirable feature. Larger
discrepancies are observed for the surface thickness (lowest panel in
Fig. \ref{densrelated}).  All forces have a tendency to underestimate
the surface thickness. That is a common feature of the RMF
models. NL-Z2 and PC--F1 included that observable in the fit and it is
then no surprise that they yield an acceptable agreement with data.
The two other forces produce to small surface thickness. The deviation
ranges up to 10\%. That is outside the range which could be explained
by possible ground state correlation effects.

\subsubsection{Density Profiles and Formfactors}

Fig. \ref{densities} shows the baryon densities, $\rho_{\rm V}(\vec{r})$
in Eq. (\ref{eq:locdens}),
for the nuclei $^{48}$Ca and $^{100}$Sn.
They all display the typical pattern of a box-like distribution with
smoothened surface and oscillations on top \cite{FRI82}. The
oscillations are an unavoidable consequence of shell structure.
$^{100}$Sn shows in addition the suppression of center density due to
the repulsive Coulomb force.
All forces produce about the same bulk properties, i.e. the overall
extension, center density, and surface profile. But there are sizeable
differences for the amplitude of the shell oscillations.  RMF-FR
produces more than factor two larger oscillations than RMF-PC (and
even that is still a bit larger than the experimentally observed
oscillations). The reason is that the finite-range folding is more
forgiving what these oscillations is concerned. It seems that the
final nuclear potential is fixed by the data to have in all cases
about the same profile with not too large oscillations. The amplitude
of oscillations in the density carries fully through to the potentials
in case of point-coupling. Thus the model needs to curb down the
initial amplitude. In finite-range models, however, the densities are
smoothened by folding with the meson propagator which gives more
leeway for oscillations on the density. Comparison with experimental
oscillations could help to decide between finite-range and zero-range
models. But just this observable of shell oscillations is heavily
modified by all sorts of ground state correlations
\cite{REI92}. These have first to be fully understood before drawing
conclusions on the range of the effective Lagrangian.\\
 
For the nucleus $^{48}$Ca, for which experimental data are
available, we compare the charge formfactor with the predictions of our
models. The experimental data are taken from Ref. \cite{Vri87}, where
the charge density is parameterized by a Fourier-Bessel series with
the coefficients determined directly from the data. This density is
then Fourier transformed to obtain the formfactor. We show it in
Fig. 11, together with the RMF predictions, in the momentum range
covered by the original analysis. Of special importance are the
first root and the height of the first maximum for finite momentum
transfer, as they
correspond to the diffraction radius and the surface thickness.
We see that all forces overestimate somewhat the first root of the
formfactor leading to a slightly too small diffraction radius. They reproduce well the following minimum which leads to an accurate prediction
of the surface thickness. Note, however, that both observables were part of
the fitting procedure for the forces PC-F1 and NL-Z2. Going to higher
momentum transfer, we see that all forces reproduce the second zero
of the formfactor and that the two RMF-PC forces agree nicely with experiment
concerning the following maximum, while the two RMF-FR forces overestimate
it somewhat. This indicates that the momentum expansion of the
RMF-PC model appears to work well in that respect up to momentun-transfer $q \approx 3.0~[{\rm fm}^{-1}]$.

\subsection{Spin-Orbit Splittings}

Figure \ref{ls_splittings} shows the relative errors for a selection
of spin-orbit splittings in  $^{16}$O, $^{132}$Sn, and
$^{208}$Pb. We have taken care to choose splittings which can be deduced
relieably from spectra of neighbouring odd nuclei
\cite{RUT98}.  All RMF
forces, except for PC-LA, perform very well.  It was shown in a former
study that RMF-FR is much better in that respect than many Skyrme
forces \cite{NLZ2}.  We see now that the well fitted point coupling
model PC--F1 does as well as RMF-FR. The ability to describe the
spin-orbit force correctly is thus a feature of the relativistic
approach.
The force PC-LA falls clearly below the others.  The poor performance
is related to the too weak fields at large densities, see
Fig. \ref{nucmatfig} and related discussion. The example demonstrates
that one needs a sufficiently large set of observables to pin down the
nuclear mean field sufficiently well.
The argument is corroborated by Fig. \ref{pb208ls} where we have a
quick glance at the effective spin-orbit potentials 
$\propto\nabla[2m_N-V_S-V_V]^{-1}$. 
The three well performing models
have all very similar potentials whereas PC-LA has a 10\% stronger
spin-orbit potential which is shifted a little bit to larger radii. This 
difference yields the observed mismatch in
the spin-orbit splittings. In turn, this figure shows that the allowed
variations on the mean fields (here the spin-orbit potential) are
rather small.

\subsection{Magic Numbers for Superheavy Nuclei}

The prediction of new magic shell closures in super-heavy elements
varies amongst the mean field models \cite{RUT97}. For protons one has
a competition between $Z$=114, 120, and 126. For neutrons one finds
$N$=172 and 184. The RMF-FR models agree in predicting a doubly magic
$^{292}120_{172}$. Precisely the same result emerges from PC--F1. This
doubly magic nucleus is thus a common feature of relativistic models.
For the density profile of $^{292}120_{172}$, we observe a central
depression in accordance with other mean-field approaches
\cite{NLZ2,DEC99,BER01}.\\

In deformed calculations done in the way as described in
Ref. \cite{BUE98}, we obtain, again in agreement with other relativistic
models, deformed shell closures at $Z=104$ for the protons and $N=162$
for the neutrons. The nuclei in that region of the nuclear chart have
deformations with $\beta_2 \approx 0.2-0.3$.  Thus also in the deformed
case, these different types of RMF models agree well concerning their
predictions of shell structure for superheavy elements.

\section{QCD Scales and Chiral Symmetry}
 
QCD is widely believed to be the underlying theory of the strong
interaction. However, a direct description of nuclear structure
properties in terms of the {\it natural} degrees of freedom of that
theory, quarks and gluons, has proven elusive. The problem is that at
sufficiently low energy, the {\it physical} degrees of freedom of
nuclei are nucleons and (intranuclear) pions.  Nevertheless, QCD can
be mapped onto the latter Hilbert space and the resulting effective
field theory is capable in principle of providing a dynamical
framework for nuclear structure calculations. This framework is
usually called chiral perturbation theory, $\chi$PT, \cite{We90}.\\
 
Two organizing principles govern this $\chi$PT: (1) (broken) chiral
symmetry (which is manifest in QCD) and (2) an expansion in powers of
$(Q/\Lambda)$, where $Q$ is a general intranuclear momentum or pion
mass and $\Lambda$ is a generic QCD large-mass scale ($\sim 1\,{\rm
GeV}$) , which in a loose sense indicates the transition region
between quark-gluon degrees of freedom and nucleon-pion degrees of
freedom.  Chiral symmetry is a direct consequence of the (approximate)
conservation of axial vector currents. 
This symmetry provides a crucial constraint in the construction of
interaction terms in the nuclear many-body Lagrangian: a general term
has the structure $\sim\;(Q/\Lambda)^{N}$ and $N\;\geq\;0$ is
mandated. Higher-order constructions in perturbation theory (loops)
will involve higher powers of $(Q/\Lambda)$ that will, consequently,
be smaller.  This mapping from {\it natural} to {\it effective}
degrees of freedom results in an infinite series of interaction terms
whose coefficients are unknown and must be determined.\\
 
In 1990, Weinberg \cite{We90} introduced $\chi$PT into nuclear physics
and showed that Lagrangians with (broken) chiral symmetry predict the
suppression of N-body forces. He accomplished this by constructing the
most general possible chiral Lagrangian involving pions and low-energy
nucleons as an infinite series of allowed derivative and contact
interaction terms and then using QCD energy (mass) scales and
dimensional power counting to categorize the terms of the series
according to $(Q/\Lambda)^{N}$.  He chose $\Lambda$ equal to the
$\rho$ meson mass of 770 MeV.  This led to a systematic suppression of
N-body forces, that is, two-nucleon forces are stronger than
three-nucleon forces, which are stronger than four-nucleon forces,
and so forth.
Thus, the infinite series of interaction terms is not physically infinite. \\
 
Following Manohar and Georgi \cite{MG84} we can scale a generic
Lagrangian term of the physical series as
\begin{equation}
{\cal L} \sim -c_{l m n}
\biggl[ \frac{\overline{\psi}\psi}{f^2_{\pi} \Lambda} \biggr]^l
\biggl[ \frac{\vec{\pi}}{f_{\pi}} \biggr]^{m}
\biggl[ \frac{\partial^{\mu}, m_{\pi}}{\Lambda} \biggr]^n
f^2_{\pi} \, \Lambda^2 \, ,
\label{V.2}
\end{equation}
where $\psi$ and $\vec\pi$ are nucleon and pion fields, respectively,
$f_{\pi}$ and $m_{\pi}$ are the pion decay constant, 92.5 MeV, and
pion mass, 139.6 MeV, respectively, $\Lambda = 770$ MeV is the $\rho$
meson mass as discussed above, and ($\partial^\mu$, $m_\pi$) signifies
either a derivative or a power of the pion mass. Dirac matrices and
isospin operators (we use $\vec{t}$ here rather than $\vec{\tau}$)
have been ignored. Chiral symmetry demands \cite{We79}
\begin{equation}
\Delta = l + n - 2 \geq 0 \: ,
\label{V.3}
\end{equation}
such that the series contains only {\it positive} powers of
(1/$\Lambda$). If the theory is {\it natural} \cite{MG84,Ly93}, the
Lagrangian should lead to dimensionless coefficients $c_{lmn}$ of
order unity. Thus, all information on scales ultimately resides in the
$c_{lmn}$. If they are natural, QCD scaling works. \\
 
An explicit pionic degree of freedom is absent in the RMF. It has been
tacitly eliminated in favor of an effective Hartree theory where the
pion effects contribute to the various effective couplings. But various
many-body effects are encompassed in the model parameters as well and
may mask the underlying chiral structure.  Nonetheless, it is
worthwhile to classify the actual RMF-PC according to naturalness.
Without pions, Eq. (\ref{V.2}) reduces to
\begin{equation}
{\cal L} \sim -c_{l n}
\left[ \frac{\overline{\psi}\psi}{f^2_{\pi} \Lambda} \right]^l
\left[ \frac{\partial^{\mu}}{\Lambda} \right]^n
f^2_{\pi} \, \Lambda^2 \,
\label{V.4}
\end{equation}
and the chiral constraint Eq. (\ref{V.3}) remains unchanged.\\
 
Our test of naturalness does not care whether a particular
$c_{lmn}\:{\rm{or}}\:c_{ln}$ coefficient has the value 0.5 or 2.0 or
some other value near 1. Changing (refining) the model by adding terms
would change {\it all} of the $c_{lmn}\:{\rm{or}}\:c_{ln}$, but the
same test of naturalness still applies. Adding new terms would simply
change a specific coefficient by an amount $\sim$ 1 (or less). That
is, testing naturalness is largely and uniquely independent of the
details, such as adding pions or performing more sophisticated nuclear
calculations, provided the framework is given by Eqs. (\ref{V.2}) -
(\ref{V.4}) while the physics is introduced via the measured
observables of nuclei.\\

The early RMF-PC parameterization of \cite{NHM92} was tested for
naturalness in \cite{FML96}.  The nine empirically fitted coupling
constants as such span 13 orders of magnitude (ignoring
dimensions).  Scaling them in accordance with the
QCD-based Lagrangian of \cite{MG84} using Eq. (\ref{V.4}), and taking into
account the role of chiral symmetry in weakening N-body forces
\cite{We90,We79} using Eq. (\ref{V.3}), yields that six of the nine scaled
coupling constants are {\it natural}.  Later work \cite{DGM97}
refitting the model using the same Lagrangian ansatz as before
resulted in two additional solutions where seven of the nine coupling
constants are natural. These results provide evidence that {\it QCD
scaling and chiral symmetry apply to finite nuclei} and, therefore, may
assist in the selection of physically admissable nuclear structure
interactions. However, one also concludes that the NHM
Lagrangian \cite{NHM92} may require more and/or different interaction terms, and
this conclusion has led to our present study.  It is important to note
that the work summarized above did not test QCD, or chiral symmetry,
but rather {\it effective Lagrangians} whose construction is {\it
constrained} by QCD and chiral symmetry.\\
 
A more extended RMF-PC adjustment was performed later \cite{RF97}. This work
also found naturalness and dimensional power counting to be extremely
useful concepts in constructing realistic chiral effective Lagrangian
expansions. Their expansions are based upon the relativistic
mean-field meson models of quantum hadrodynamics (QHD)
\cite{SW86,FST97}.  Thus, each term in their Lagrangian corresponds to
the leading-order expansion of one appearing in an appropriate
\cite{FST97} QHD-based meson-nucleon Lagrangian.
   Accordingly, their
   RMF-PC Lagrangian contains nucleon densities of isoscalar-scalar,
   -vector, -tensor and isovector-vector, -tensor character, with each
   tensor term appearing only as a product with its corresponding vector
   term. No isovector-scalar terms appear due to their absence in the
various QHD approaches. In their fourth-order truncation,
the best-fit set ( 16 coupling constants, unconstrained search)
contained fourteen natural and two unnatural coupling constants,
whereas the worst-fit set (14 coupling constants, constrained search)
is the one set containing {\it all} natural coupling constants.
Note, however, that the coupling constants of the derivative terms
were constrained by the appropriate meson masses, as described in Sec.
\ref{sec:preview}.
Nevertheless, their study concludes that {\it naturalness and
dimensional power counting are compatible with and implied by the
measured ground-state properties of finite nuclei}.\\
 
We now turn to these same considerations for the sets of coupling
constants determined in our present study that are tabulated in
Sec. III.  Applying Eqs. (\ref{V.3}) and (\ref{V.4}) to the sets of
dimensioned coupling constants in Tables \ref{p-f1} through \ref{p-f4}, 
and using
Weinberg's \cite{We90} choice of the $\rho$ meson mass (770 MeV) for
the QCD large-mass scale $\Lambda$, we obtain the corresponding sets
of QCD-scaled coupling constants listed in Table \ref{V.T1}, together
with the additional information of expansion order in $\Lambda$, number
of coupling constants, number of natural coupling constants amongst them, and
ratio of maximum and minimum scaled coupling constants in the set.
The table also shows the $\chi^2$ per degree of freedom. The sets are
ordered according to increasing values of this quantity.
For our present work we require a more quantitative definition of a
{natural set} of coupling constants than the various
interpretations of the usual phrase {``of order one''} which have
been applied \cite{FST97,FR98,FML96,DGM97}:
a set of QCD-scaled coupling constants is {\it natural} if their absolute
values are distributed about the value 1 {\it and} the ratio of the
maximum value to the minimum value is {\it less than 10}.
   We now discuss each set of
   QCD-scaled coupling constants appearing in Table \ref{V.T1}.
 
\subsection{Interaction PC--F1}
 
The PC--F1 interaction is the most physically realistic interaction
that we have found. It reproduces the measured observables used to
determine its coupling constants more exactly than any of our
other interactions, as can be seen by inspection of the
$\chi^{2}_{\rm df}$ values in Table \ref{V.T1}. Its predictive power is
also better than that of the other interactions as has been shown in
Sect. IV. The nine QCD-scaled coupling constants are all natural and
the ratio of the maximum to the minimum is 8.92, thus satisfying our
definition of a natural set of QCD-scaled coupling constants. So far
as we are aware, this is the {\it first complete set of natural
QCD-scaled coupling constants}, with order up to $\Lambda^{-2}$, that
has been obtained with unconstrained least-squares parameter
adjustment to measured ground-state observables.

\subsection{Interaction PC--F2}
 
The form of the PC--F2 interaction is identical to that of PC--F1
except for the addition of the isovector-scalar term in Eq.
(\ref{eq:lagrang}). The most likely corresponding isovector-scalar meson
is the $\delta$ meson with a mass of 983 MeV and a relatively weak
coupling constant, $g_{\delta} \sim 2$, according to Machleidt
\cite{M86}. Thus, its contribution is expected to be small.
Nevertheless, the QCD-scaled coupling constant should be of order 1 if
$\delta$ meson exchange has a physical role in the strong interaction
occurring in finite nuclei in the ground state. We will return to this
topic in our discussion of the PC--F4 interaction.  Nine of the ten
QCD-scaled coupling constants of this interaction are natural whereas
that of the isovector-scalar term, $\alpha_{\rm TS} = 2.34 \cdot 10^{-6}~{\rm MeV}^{-2}$, is very
small and unnatural, as one would expect from the above discussion. 
This small value is responsible for the relatively large ratio of
67.5 leading to the conclusion that this QCD-scaled set of coupling
constants is not natural. This deviation from naturalness (here
and for the following two forces) can have several reasons. There may
be a yet undiscovered symmetry or the minimization procedure has 
found only a local minimum.
 
\subsection{Interaction PC--F3}
 
The form of the PC--F3 interaction is also identical to that of PC--F1
except for the addition of the quartic isovector-vector term, Eq. 
(\ref{eq:pcf2}). 
This was done in hopes of
producing a sign change in either of the two other isovector-vector
   terms, $\alpha_{\rm TV}$ or $\delta_{\rm TV}$, so that their ratio would be
   positive, thus satisfying expectations based upon the 1st order
   expansion of the propagator for the $\rho$ meson, as discussed in
Sec. \ref{sec:preview}.
   The sign change, however, did not occur.
Again, nine of the ten QCD-scaled coupling constants of this interaction are
natural whereas that of the quartic isovector-vector term, $\gamma_{\rm TV}
= -5.4 \cdot 10^{-17}~{\rm MeV}^{-8}$, is very large and unnatural. 
This large value is responsible for the very large ratio of 264.7,
again leading to the conclusion that this QCD-scaled set of
coupling constants is not natural.

\subsection{Interaction PC--F4}
 
The PC--F4 interaction is built from the PC--F1 interaction by the
addition of isovector-scalar terms that are quadratic and derivative
of quadratic in the corresponding density. This continues the attempt
with the PC--F2 interaction to address the role of the $\delta$ meson
by including both terms that are necessary to simulate the
propagator. While only nine of the eleven QCD-scaled coupling
constants are natural, and $|{\rm max}|/|{\rm min}|$ is a factor $\sim$ 10 worse than
that of the PC--F1 interaction, it is very interesting to observe that
the signs of the two new terms are identical and thus they correctly
simulate the expansion of the propagator for the $\delta$ meson. Not
only that, but the corresponding signs for the $\rho$ meson are, for
the first time in the present study, also identical.  Thus, the
expansions of the propagators for the two isovector mesons appearing
in the PC--F4 interaction have the correct relative
signs.
Nevertheless, the maximum ratio is yet large, 103.8, leading again
to the conclusion that this QCD-scaled set of coupling constants
is not natural.
We believe, however, that the PC--F4
interaction should be studied further. \\ 
 
We conclude this section by noting that the PC--F1 interaction is
one that leads to a physically admissable Lagrangian from the
simultaneous points of view of (a) predictability and (b)
naturalness. We have therefore demonstrated that QCD scaling
and chiral symmetry apply to finite nuclei.
 
\section{CONCLUSIONS}

We have investigated the properties and applicability of a relativistic point-coupling model for nuclear structure calculations. 
To answer the question whether the point-coupling model can reach a predictive power comparable to other state-of-the-art mean field approaches, like the RMF-FR and SHF models, we have carefully performed a $\chi^2$ minimization
combining two different search algorithms, and have been guided by
expectations of naturalness in physically realistic extracted coupling
constants.
The resulting set of coupling constants is PC--F1 in Table \ref{p-f1}.
It has been used to test the predictive power of the RMF-PC in a variety
of applications ranging from saturated symmetric nuclear and neutron
matter, binding energies in isotopic and isotonic chains to formfactor-
and shell-structure related observables (rms charge radii, diffraction
radii, surface thicknesses and spin-orbit splittings) and the fission barrier of $^{240}$Pu.
The net result is that the RMF-PC model with PC--F1 actually has reached
the quality of competing approaches. In some of these comparisons we
discovered the influence of finite versus zero range in the models. For
example, the density profiles of RMF-PC are generally smoother than those
with RMF-FR.
Like the SHF model, the point-coupling model naturally leads to a rather high incompressibility in nuclear matter, $K = 260~{\rm MeV}$. 
And like the established RMF-FR forces, the point-coupling exhibits some unresolved isovector trends and a rather high symmetry energy in nuclear matter.
The model performs well in deformed calculations. Also, the spin-orbit splittings are reproduced in a manner comparable to the finite-range models, showing
that the relativistic framework is important here rather than the
finite range. \\

Attempts to extend the effective Lagrangians utilizing additional
isovector terms proved to be elusive: the additional coupling constants
can only be loosely determined with the existing set of experimental
observables. Thus the problem remains the same as in RMF-FR and SHF
approaches, namely, that the experimental observables are very highly
correlated with respect to the values of the coupling constants. This
means that highly accurate experimental observables corresponding
to large isospin are required to determine the isovector properties
of the model more completely. \\

We have been guided by naturalness in the extraction of our sets
of coupling constants and have found that those of the set
PC--F1 are all natural. In fact, so far as we are aware, PC--F1 is
the first complete set of natural coupling constants that have been
determined in an unconstrained search.
This result, together with the predictability of PC--F1, demonstrates
that QCD scaling and chiral symmetry apply to finite nuclei.
It appears, from the sets PC-F2 and PC-F4, that either $\delta$-meson
exchange is not natural and is not required for a viable description
of the strong interaction in finite nuclei, or there exists an as yet
undiscovered symmetry.
We think, however, that the PC--F4 interaction requires further study
including possible extensions beyond eleven coupling constants
(especially following new measurements on high-isospin nuclei)
because the extracted isovector coupling constants all have the
right signs to satisfy expectations from the expansions of their
propagators. \\

The point-coupling model discussed here may be viewed as a {\em missing
link} between the established SHF and RMF-FR models. With it, one can
separately investigate the influence of finite range versus zero range
and relativistic framework versus nonrelativistic framework. This is
important because, as we have learned, there are differences in the
predictions from the two model classes which cannot easily be mapped
onto the separate features of the two classes. We believe that
future work should include more detailed studies of the isovector
components of the relativistic effective Lagrangians and, perhaps
more importantly, the influence of the Fock terms via the Fierz
relations. Systematic studies of relativistic Hartree-Fock calculations
using RMF-PC will provide further linkages, on the one hand, with
relativistic Hartree calculations using RMF-FR, and on the other hand,
with nonrelativistic Hartree-Fock calculations using SHF. Work in these
directions is in progress.

\section{ACKNOWLEDGEMENTS}
The authors would like to thank M.~Bender, J.~Friar, W.~Greiner,
P.~M{\"o}ller, A. J. Sierk and A. Sulaksono for many valuable discussions.  This work was supported
in part by the Bundesministerium f\"ur Bildung und Forschung (BMBF),
Project No.\ 06 ER 808, by Gesellschaft f\"ur Schwerionenforschung
(GSI), and the U. S. Department of Energy.

%bibliography

% end bib

% figure captions -----------------------------------------------------
%----------------------------------------------------------------------

\begin{figure}[h]
\centerline{\epsfxsize=15cm \epsfbox{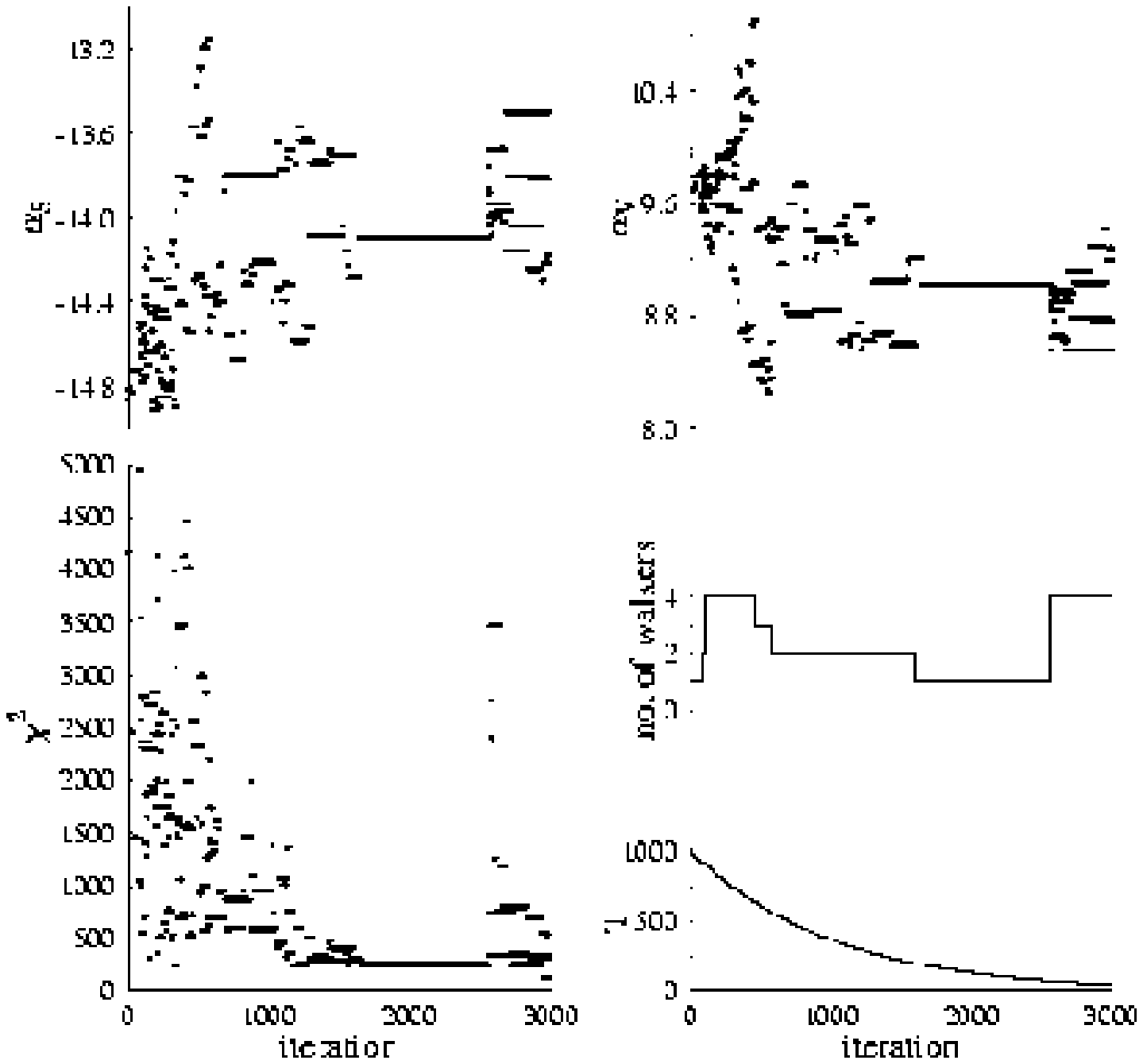}}
\caption{
The upper two figures show the positions of the walkers in parameter
space in $\alpha_{\rm S}$ and $\alpha_{\rm V}$ direction. For each
walker, a dot is printed. The figure below on the left side shows the
corresponding $\chi^2$ values. The other two figures indicate the size
of the population and the external temperature.
}\label{mcrun}
\end{figure}

\begin{figure}[htb]
\epsfxsize=12cm
\epsfbox{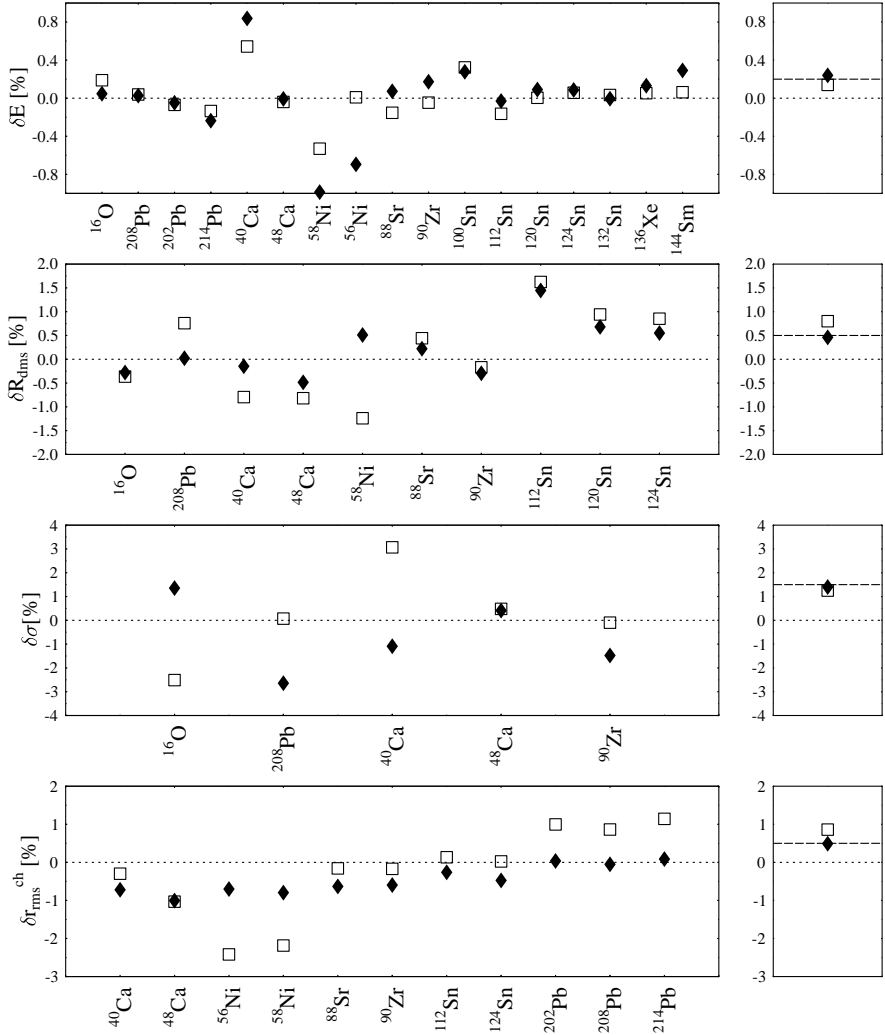}
\caption{
Errors in percent for the observables binding energy, diffraction
radius, surface thickness and rms charge radius for PC--F1 (filled
diamonds) and NL--Z2 (open squares) are seen on the left. The right panels 
show the absolute mean errors
for the corresponding observables, where the dashed lines indicate the chosen relative errors $\Delta O$ in the fitting procedure.
}\label{fitnuclei}
\end{figure}

\begin{figure}[htb]
\centerline{\epsfxsize=13cm \epsfbox{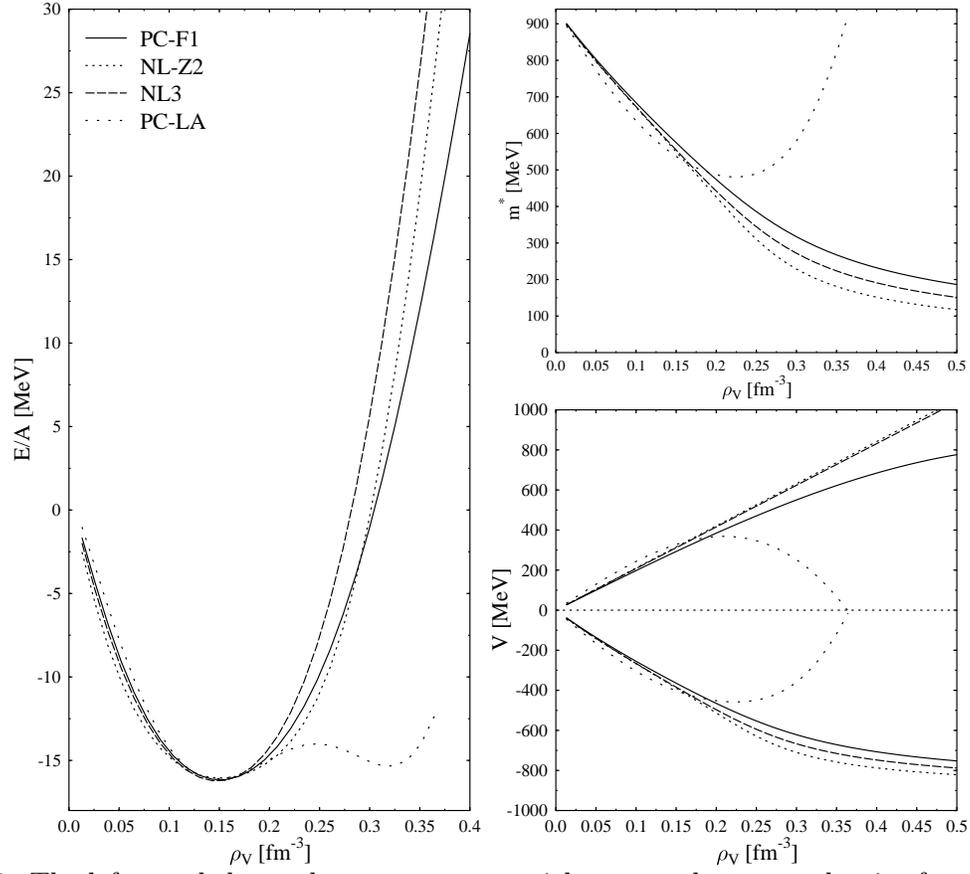}}
\caption{The left panel shows the energy per particle versus the vector 
density for the forces as indicated. On the right side, the effective 
nucleon mass (upper figure) and the scalar and vector mean-field potentials 
(lower figure) are drawn as emerging from the calculations.}
\label{nucmatfig}
\end{figure}

\begin{figure}[htb]
\centerline{\epsfxsize=13cm \epsfbox{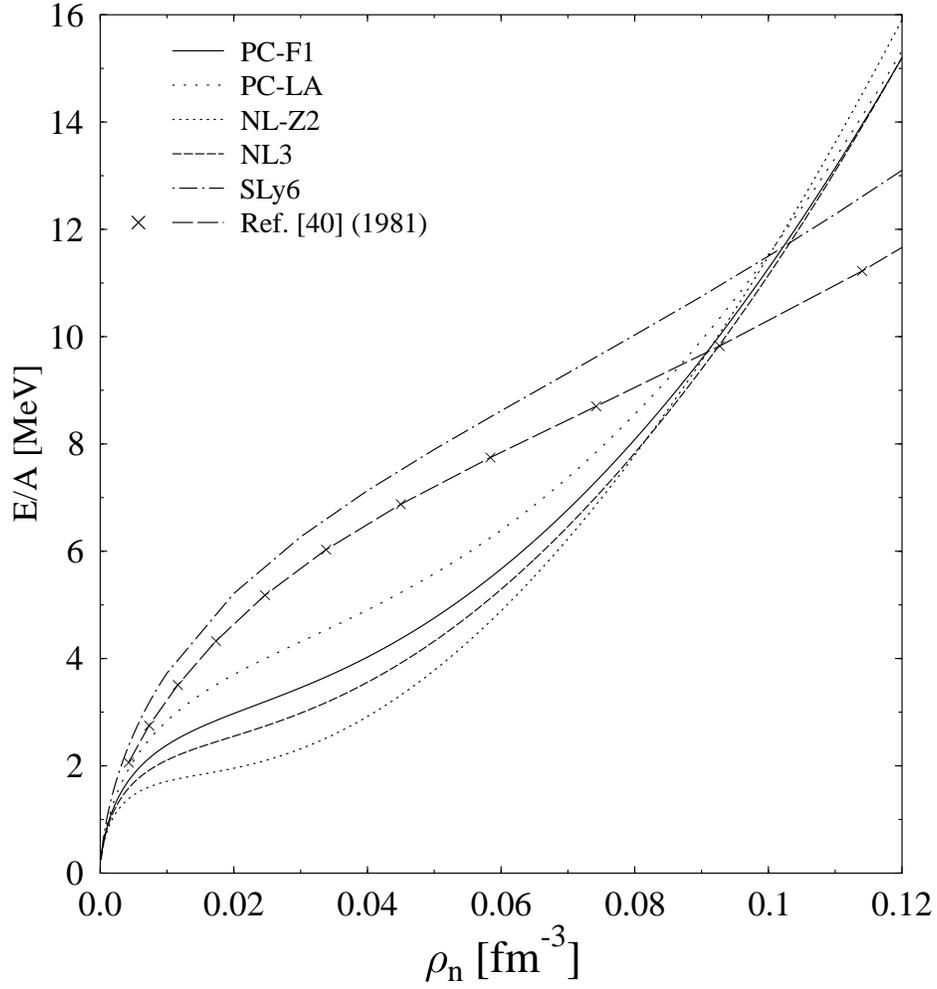}}
\caption{Energy per particle versus neutron density for four RMF forces and 
the Skyrme force SLy6. The crosses mark data from Ref. \protect\cite{FRI81}.}
\label{neutronmatter}
\end{figure}

\begin{figure}[h]
\centerline{\epsfxsize=15cm \epsfbox{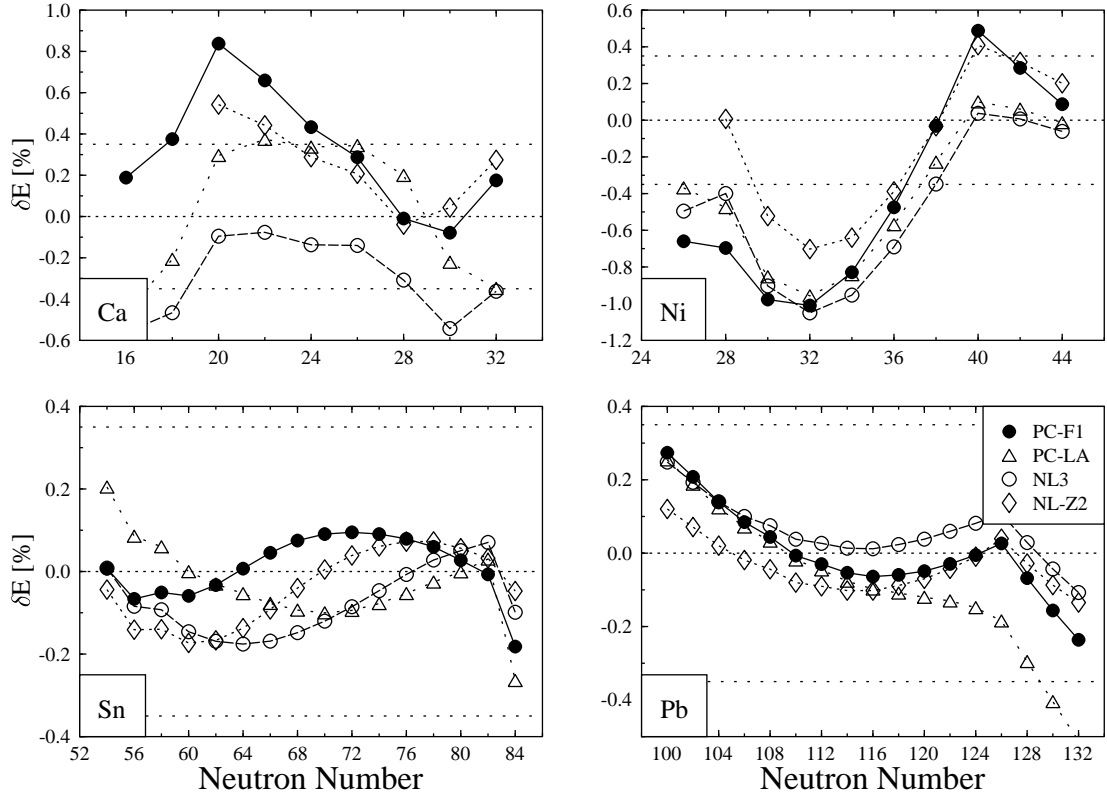}}
\caption{Deviation in \% of the calculated energies from the
experimental values in spherical calculations of isotopic chains. Note
that the scales are different for each figure. The dotted lines indicate the 
accuracy that can be demanded from the models. The experimental errors are smaller than the size of the symbols used in the figure.
}\label{isotopes}
\end{figure}

\begin{figure}[htb]
\centerline{\epsfxsize=15cm \epsfbox{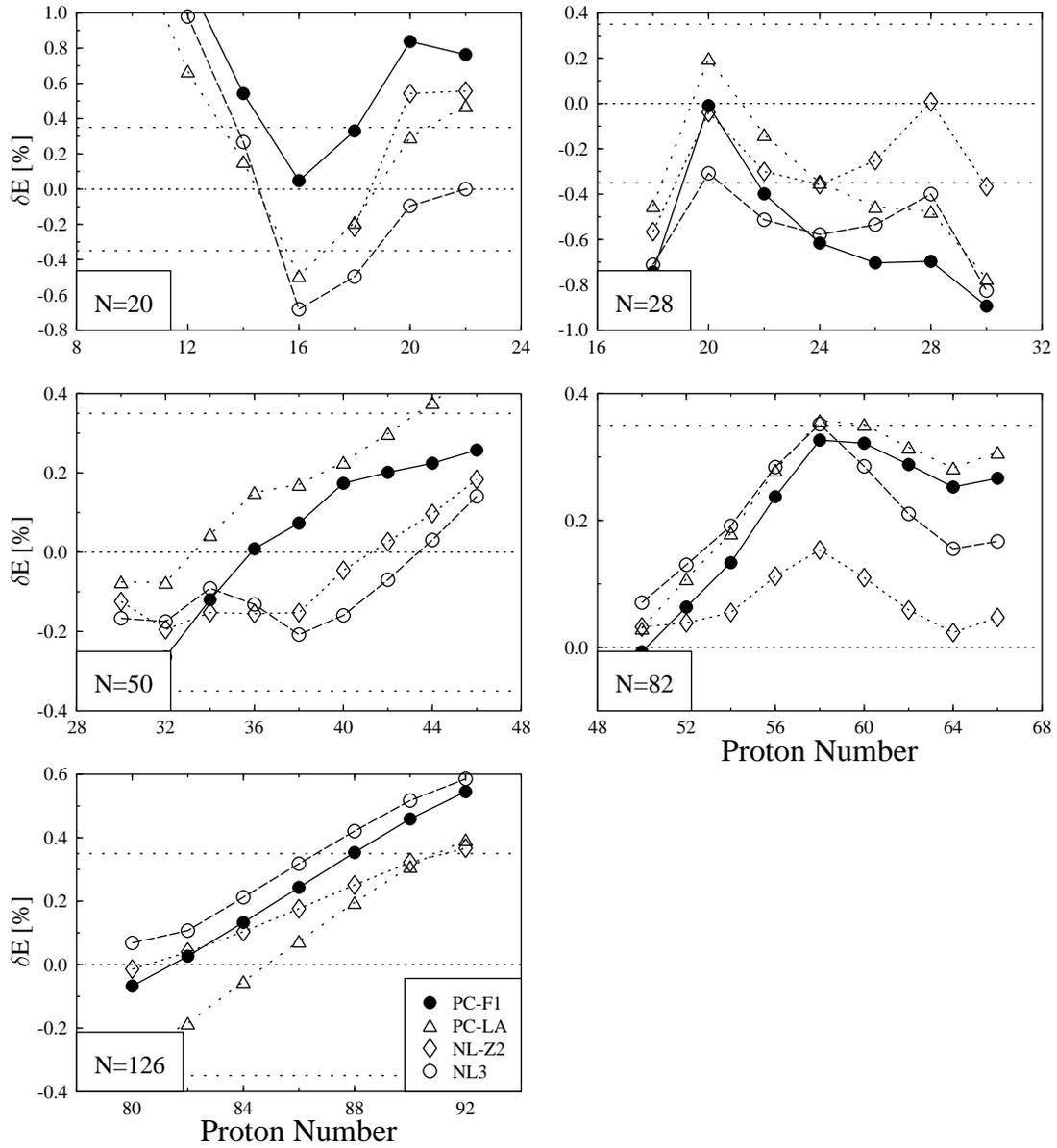}}
\caption{Deviation in \% of the calculated energies from the
experimental values in spherical calculations of isotonic chains. Note
that the scales are different for each figure. The dotted lines indicate the 
accuracy demanded for energies. The experimental errors are smaller than the size of the symbols used in this figure.}\label{isotones}
\end{figure}

\begin{figure}[htb]
\centerline{\epsfxsize=15cm \epsfbox{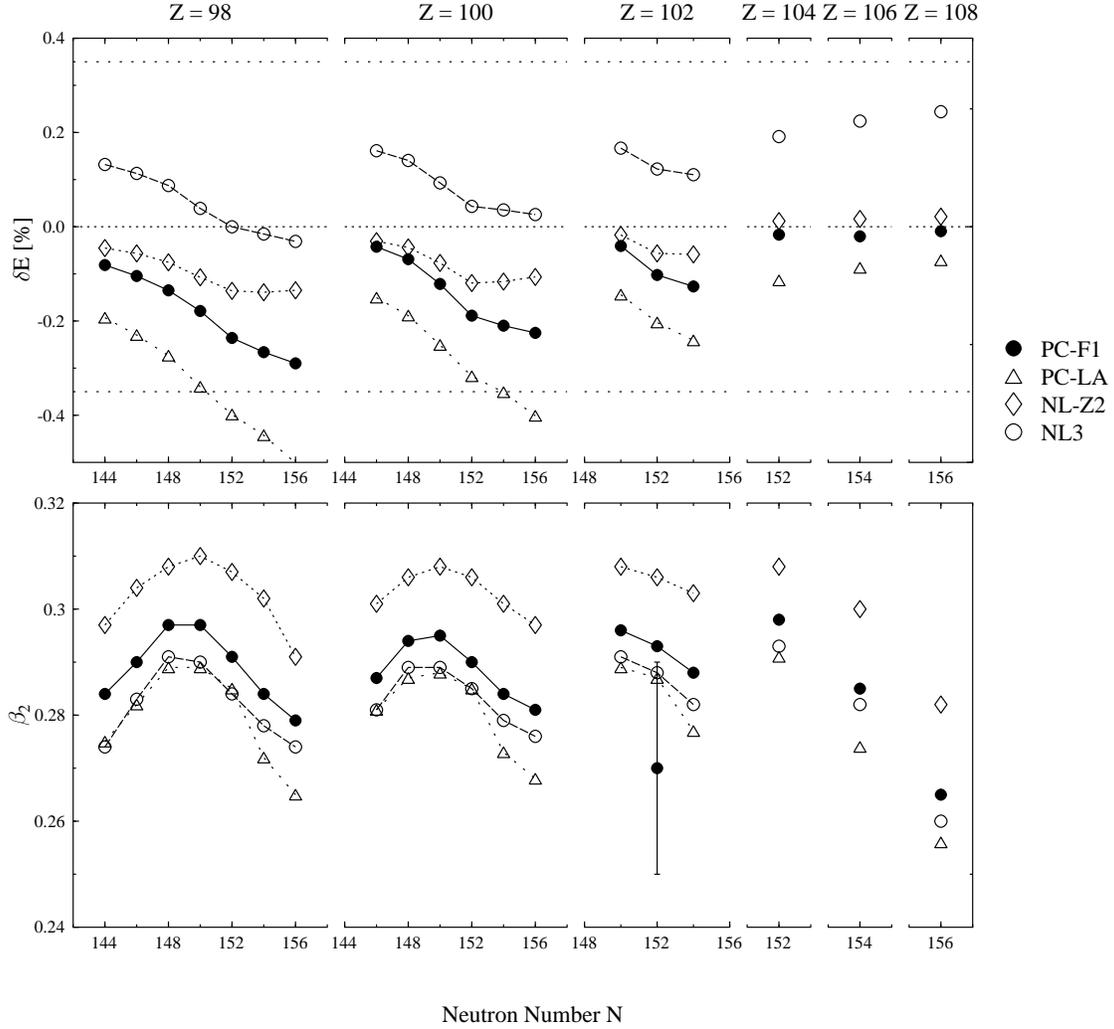}}
\caption{Deviation in \% of the calculated energies from the
experimental values (upper figure) and ground-state deformations (lower
figure) in axially deformed and reflection symmetric calculations.
The errors for the binding energies are smaller than the size of the symbols
used in this figure. The 
symbol with error bars indicates the measured ground-state deformation together with its uncertainty of $^{254}{\rm No}$ 
\protect\cite{REI99,LEI99}.}\label{heaviestgg}
\end{figure}

\begin{figure}[htb]
\centerline{\epsfxsize=15cm \epsfbox{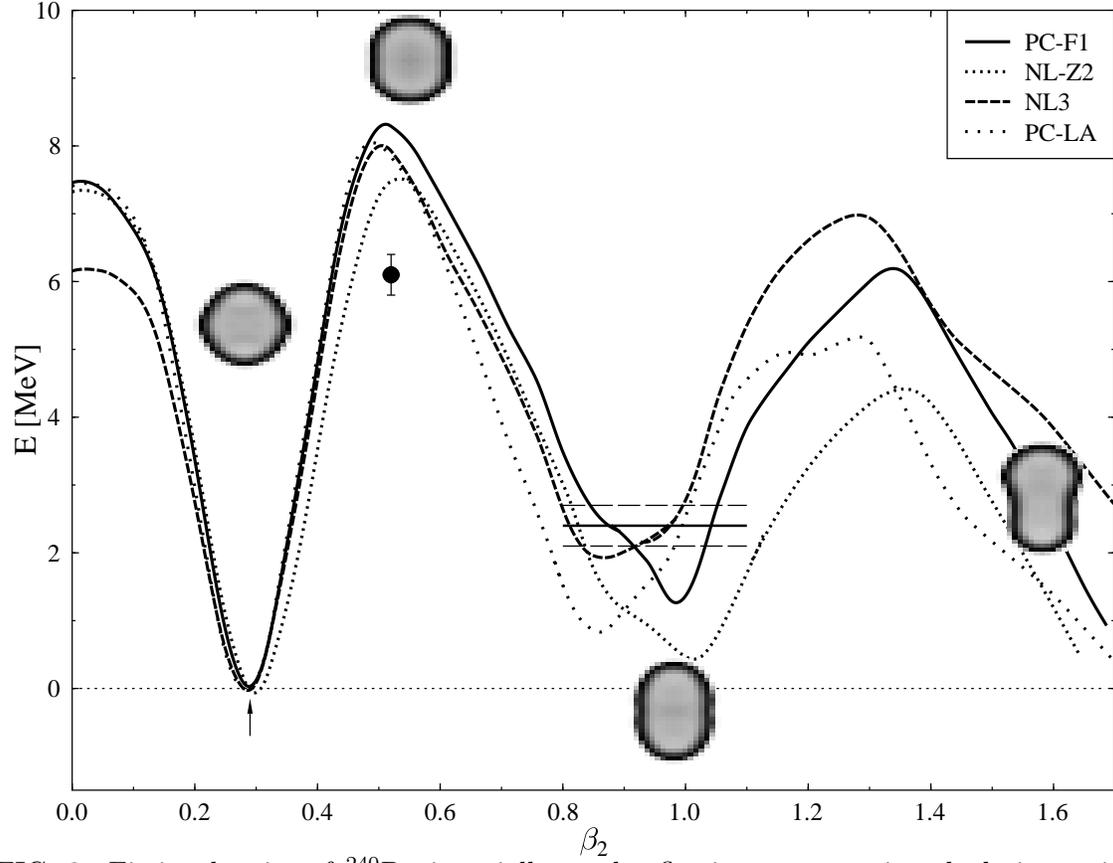}}
\caption{Fission barrier of $^{240}$Pu in axially- and
reflection-asymmetric calculations with the forces as indicated. The
experimental values for the ground-state deformation, the barrier
height, and the energy of the second minimum are indicated, respectively, with an
arrow, a symbol with error bars and three lines indicating the value and its errors. The data are taken
from Ref. \protect\cite{BJO80,FIR96,LOB70,BEM73}.}\label{pu240}
\end{figure}

\begin{figure}[p]
\centerline{\epsfxsize=14cm \epsfbox{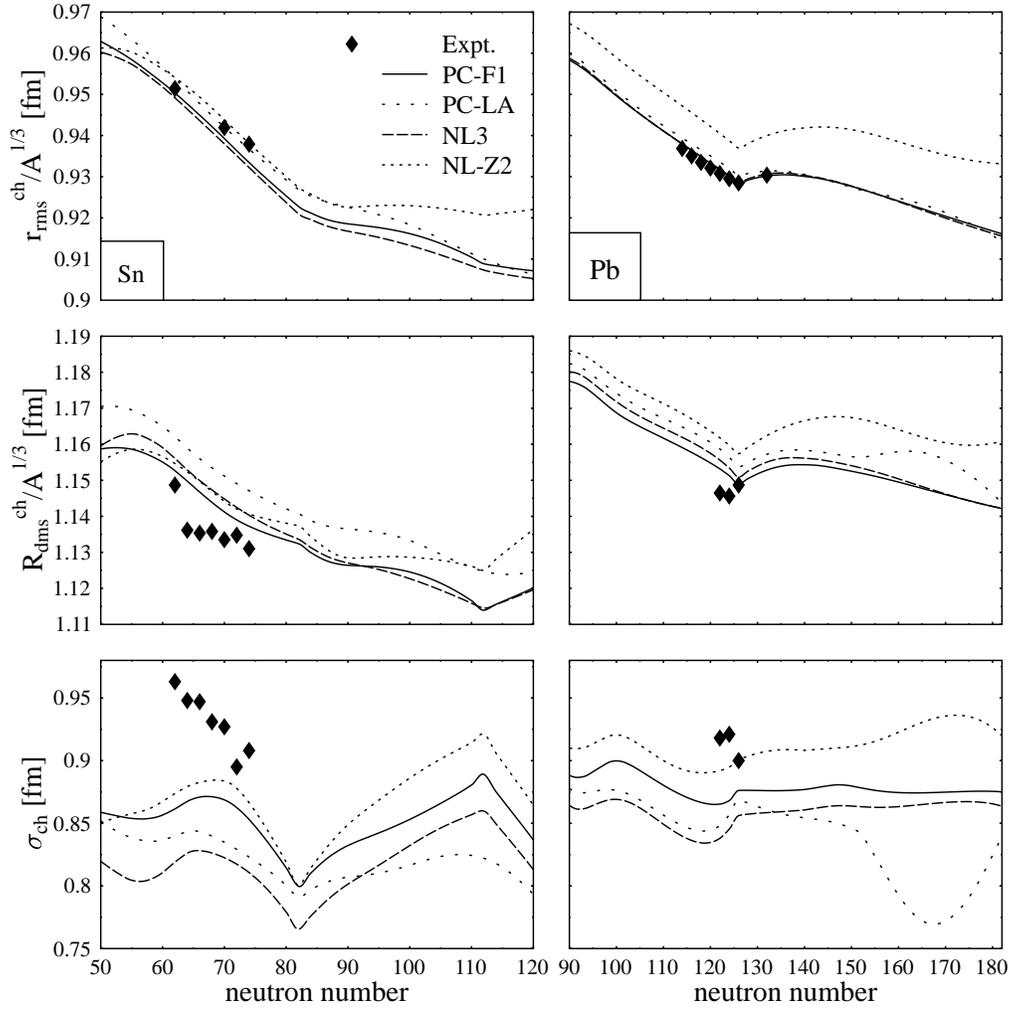}}
\caption{Surface thicknesses (lower figure), diffraction radii
(middle) and rms charge radii (upper figure) for Sn (left) and Pb
(right) isotopes emerging from spherical calculations. Note that the
radii have been divided by $A^{1/3}$ to eliminate the liquid drop
trend with mass number. The experimental data are from \protect
\cite{FRI82,OTT89,FRI95}. Their errors are smaller than
the symbols used in this figure.}\label{densrelated}
\end{figure}

\begin{figure}[thb]
\begin{minipage}[t]{7cm}
\centerline{\epsfxsize=7cm \epsfbox{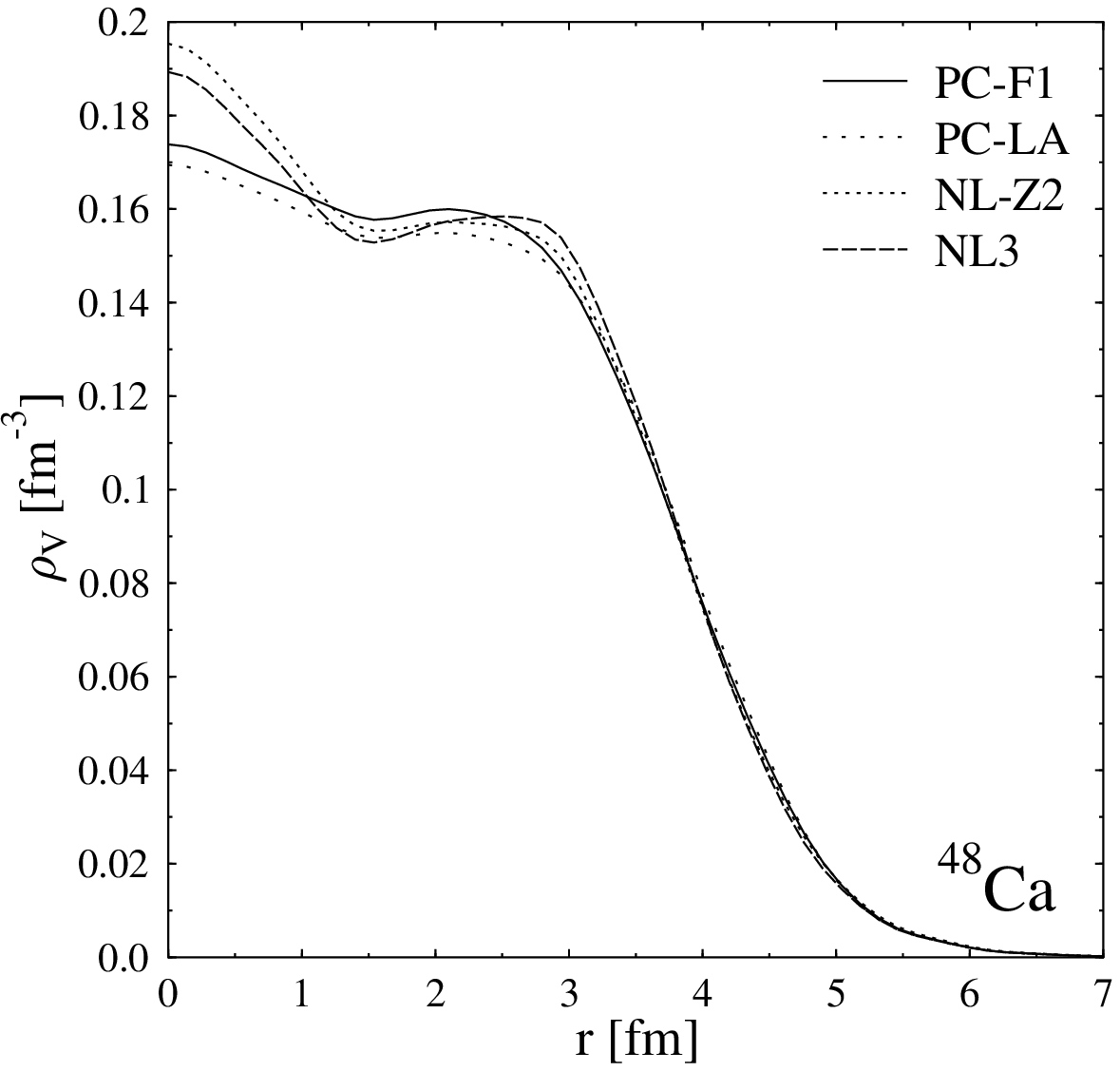}}
\end{minipage}
\begin{minipage}[b]{7cm}
\centerline{\epsfxsize=7cm \epsfbox{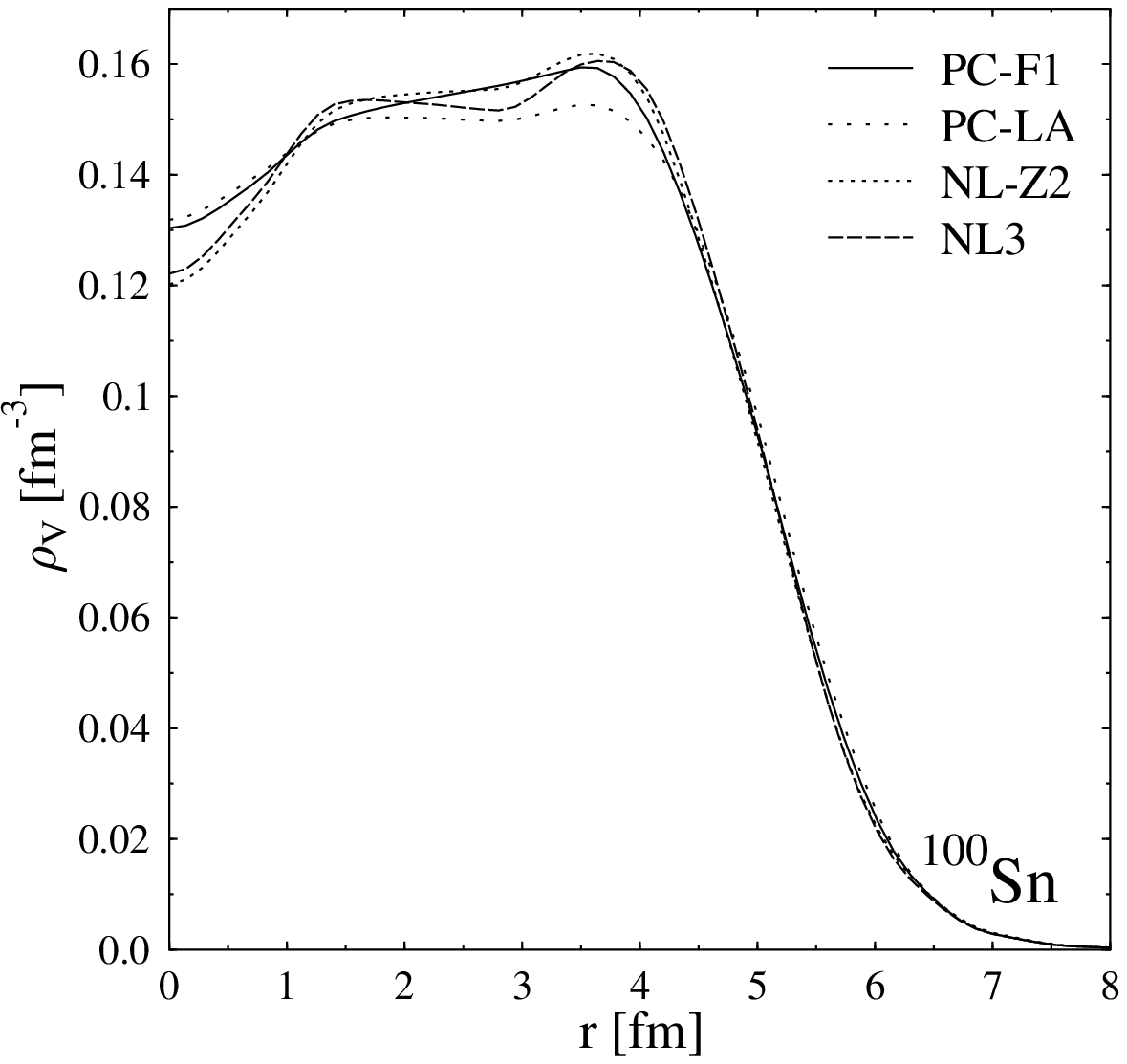}}
\end{minipage}
\caption{Total baryon densities of the nuclei
$^{48}$Ca and $^{100}$Sn emerging from spherical calculations for the four 
RMF forces under
consideration.}\label{densities}
\end{figure}

\begin{figure}[htb]
\centerline{\epsfxsize=10cm \epsfbox{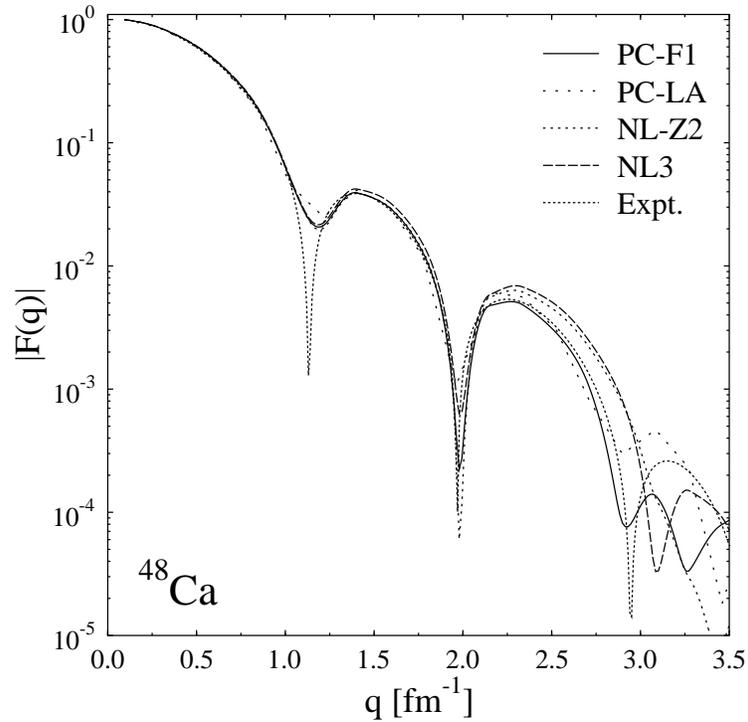}}
\caption{Formfactor of the charge density of the nucleus $^{48}$Ca for the four RMF forces under consideration. The experimental data, which are taken from Ref. \protect\cite{Vri87}, are plotted in the momentum-transfer range $q =
0.35-3.55~ [{\rm fm^{-1}}]$, as in the original analysis.}
\label{formfactor}
\end{figure}

\begin{figure}[htb]
\centerline{\epsfxsize=14cm \epsfbox{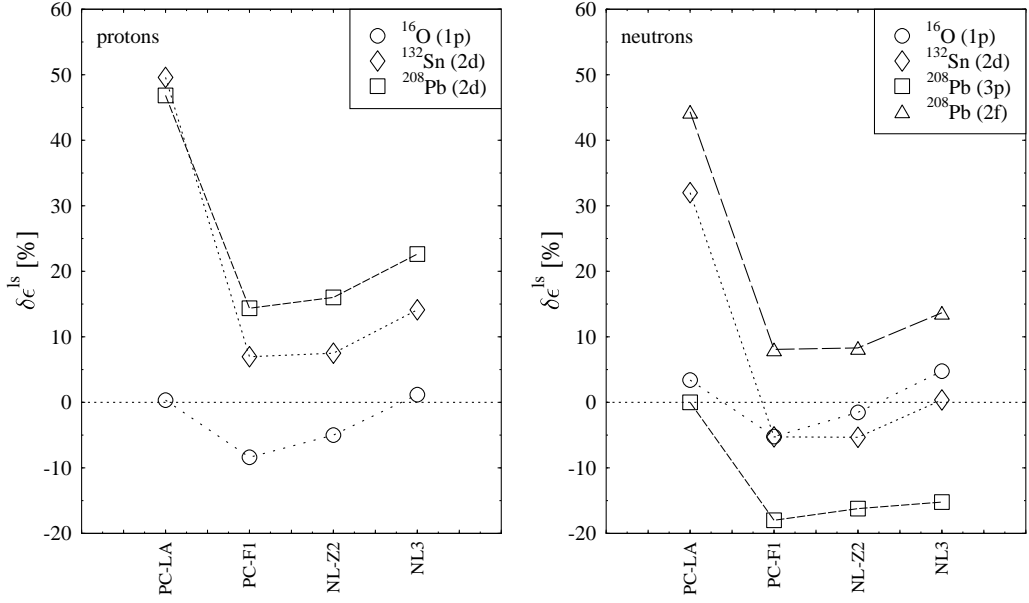}}
\caption{The percentage error in $ls$-splittings for protons (left) and for 
neutrons (right). The experimental errors are smaller than the size of the symbols used in these figures.
The lines serve to guide the eye.}
\label{ls_splittings}
\end{figure}

\begin{figure}[htb]
\centerline{\epsfxsize=7cm \epsfbox{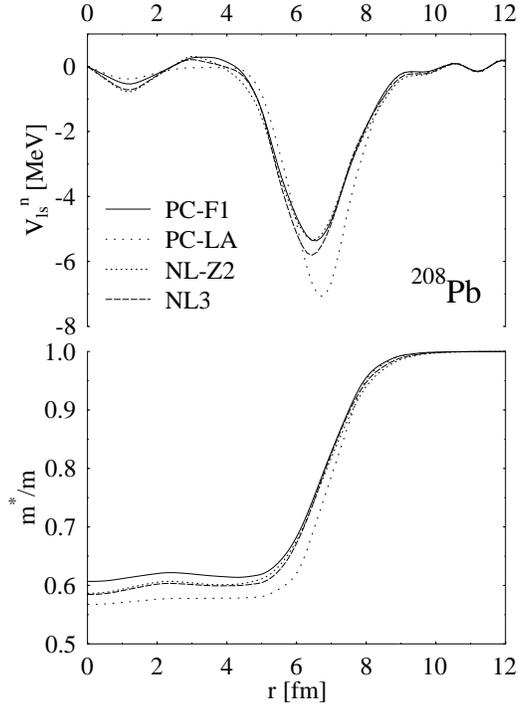}}
\caption{The neutron spin-orbit potential (upper figure) and effective
mass (lower figure) in $^{208}$Pb for the forces under
discussion. }\label{pb208ls}
\end{figure}

% tables including captions ------------------------------------------------
%---------------------------------------------------------------------------

\begin{table}[htb]
\begin{tabular}{|l|r|r|r|r|}
observable &  error & $^{16}$O & $^{88}$Sr & $^{208}$Pb \\
\hline
$E_{\rm B}$ & 0.15~\% & + & + & + \\
$r_{\rm rms}^{\rm ch}$ & 0.2~\% & + & + & + \\
$\epsilon_{\rm ls}^{\rm p}$ & 5.0~\% & + & + & + \\
$\epsilon_{\rm ls}^{\rm n}$ & 5.0~\% & + & + & + \\
\end{tabular}
\caption{
Observables and chosen errors $\Delta O$ for set 1 of nuclei used for
the SA fitting procedure.  $E_{\rm B}$ denotes the binding energy, $r_{\rm
rms}^{\rm ch}$ the rms charge radius and $\epsilon_{\rm ls}^{\rm p}$ and $\epsilon_{\rm
ls}^{\rm n}$ the proton and neutron spin-orbit splittings of one
selected level.  A $+$ indicates an observable contributing to the
total $\chi^2$. For the experimental values see Ref. \protect\cite{NHM92}.}
\label{nuc-set-1}
\end{table}

\begin{table}[htb]
\begin{tabular}{@{\hspace{0.1cm}}l@{\hspace{0.2cm}}c@{\hspace{3.0mm}}c@
{\hspace{3.0mm}}c@{\hspace{3.0mm}}c@{\hspace{3.0mm}}c@{\hspace{3.0mm}}c@
{\hspace{3.0mm}}c@{\hspace{3.0mm}}c@{\hspace{3.0mm}}c@{\hspace{3.0mm}}c@
{\hspace{3.0mm}}c@{\hspace{3.0mm}}c@{\hspace{3.0mm}}c@{\hspace{3.0mm}}c@
{\hspace{3.0mm}}c@{\hspace{3.0mm}}c@{\hspace{3.0mm}}c@{\hspace{3.0mm}}c@
{\hspace{0.1cm}}}
observable & error &  \begin{sideways}$^{16}$O\end{sideways}  &  
\begin{sideways}$^{40}$Ca\end{sideways}  &  \begin{sideways}$^{48}$Ca
\end{sideways}  &  \begin{sideways}$^{56}$Ni\end{sideways}  & \begin{sideways} 
$^{58}$Ni\end{sideways}  & \begin{sideways} $^{88}$Sr\end{sideways}  &  
\begin{sideways}$^{90}$Zr\end{sideways}  &  \begin{sideways}$^{100}$Sn
\end{sideways}  &  \begin{sideways}$^{112}$Sn\end{sideways}  & 
\begin{sideways} $^{120}$Sn\end{sideways}  & \begin{sideways} $^{124}$Sn
\end{sideways}  &  \begin{sideways}$^{132}$Sn\end{sideways}  &  
\begin{sideways}$^{136}$Xe\end{sideways}  & \begin{sideways} $^{144}$Sm
\end{sideways}  & \begin{sideways} $^{202}$Pb \end{sideways}  &  
\begin{sideways}$^{208}$Pb\end{sideways}  & \begin{sideways} $^{214}$Pb
\end{sideways}  \\ \hline
$E_{\rm B}$ & 0.2~\% &+&+&+&+&+&+&+&+&+&+&+&+&+&+&--&+&+\\
$R_{\mbox{\scriptsize dms}}$ & 0.5~\% &+&+&+&--&+&+&+&--&+&+&+&--&--&--&--&+&--\\
$\sigma$ & 1.5~\% &+&+&+&--&--&--&+&--&--&--&--&--&--&--&--&+&--\\
$r_{\mbox{\scriptsize rms}}^{\mbox{\scriptsize ch}}$ & 0.5~\% &--&+&+&+&+&+&+&
--&+&--&+&--&--&--&+&+&+\\ \hline
$\Delta_{\mbox{ \scriptsize p}}$& 0.05~MeV & - & - & - & - & - & - & - & - & - 
& - & - & - &
+ & + & - & - & -\\
$\Delta_{\mbox{ \scriptsize n}}$& 0.05~MeV & - & - & - & - & - & - & - & - 
& + & + & + & - &
- & - & - & - & -\\
\end{tabular}
\caption{
Observables and chosen errors $\Delta O$ for set 2 of nuclei used for
the Bevington fitting procedure.  $R_{\rm diffr.}$ denotes the
diffraction-radius, $\sigma$ the surface thickness, and $\Delta_{\rm
p}$ and $\Delta_{\rm n}$ are the proton and neutron pairing gaps.  A
$+$ indicates an observable contributing to the total $\chi^2$. For
the experimental values see Ref. \protect\cite{NLZ2}
}\label{nuc-set-2}
\end{table}

\begin{table}[htb]
\begin{tabular}{|c|c|c|c|c|} 
coupling constant &  value & dimension & uncorr. error [\%] & corr. 
error [\%]\\
\hline
$\alpha_{\rm S}$    &  $-3.83577\cdot 10^{-4}$ & MeV$^{-2}$ & $2.6\cdot10^{-3}$  & 
$8.3\cdot 10^{-1}$\\
$\beta_{\rm S}$     & $7.68567\cdot 10^{-11}$ & MeV$^{-5}$  & $2.1\cdot 10^{-2}$ & 
$4.7\cdot 10^{0}$ \\
$\gamma_{\rm S}$    & $-2.90443\cdot 10^{-17}$ & MeV$^{-8}$ & $7.2\cdot 10^{-2}$ & 
$1.4\cdot 10^{1}$\\
$\delta_{\rm S}$    & $-4.1853\cdot10^{-10}$ & MeV$^{-4}$ & $2.0\cdot 10^{-1}$ & 
$2.3\cdot 10^{1}$ \\
$\alpha_{\rm V}$    &  $2.59333\cdot 10^{-4}$& MeV$^{-2}$ & $3.7\cdot 10^{-3}$ & 
$1.2\cdot 10^{0}$\\
$\gamma_{\rm V}$    & $-3.879\cdot 10^{-18}$ & MeV$^{-8}$  & $4.4\cdot 10^{-1}$ & 
$5.4\cdot 10^{1}$\\
$\delta_{\rm V}$    & $-1.1921\cdot 10^{-10}$ & MeV$^{-4}$ & $6.0\cdot 10^{-1}$ & 
$7.6\cdot 10^{1}$\\
$\alpha_{\rm TV}$    & $3.4677\cdot 10^{-5}$ & MeV$^{-2}$ & $1.2\cdot 10^{0}$ & 
$1.1\cdot 10^{1}$\\
$\delta_{\rm TV}$    & $-4.2 \cdot 10^{-11}$ & MeV$^{-4}$ & $6.0\cdot 10^{1}$ & 
$1.7\cdot10^{3}$ \\
\hline\hline
$V_{\rm P}$ & $-321$ & MeV~fm$^3$ & $1.3\cdot 10^0$ & $2.0\cdot 10^0$\\
$V_{\rm N}$ & $-308$ & MeV~fm$^3$ & $1.2\cdot 10^0$ &  $2.3\cdot 10^0$\\
\end{tabular}
\caption{ The set PC--F1 of coupling constants resulting from the
final fitting procedure. In columns 4 and 5 the uncorrelated and
correlated errors are shown as originating from the fitting
procedure.
Note that the values for the pairing strengths have been
rounded according to the error marigins. The total $\chi^2$, $\chi^2$ per 
point and $\chi^2$ per degree of freedom are $\chi^2_{\rm tot} = 99.1$, 
$\chi^2_{\rm pt} = 2.11$, $\chi^2_{\rm df} = 2.75$. }\label{p-f1}
\end{table}

\begin{table}[htb]
\begin{tabular}{|c|c|c|c|c|} 
coupling constant &  value & dimension & uncorr. error [\%] & corr. 
error [\%]\\
\hline
$\alpha_{\rm S}$    & $-3.835821\cdot 10^{-4}$  & MeV$^{-2}$ & $1.9\cdot 10^{-3}$  & 
$8.1\cdot 10^{-1}$ \\
$\beta_{\rm S}$     & $7.6835\cdot 10^{-11}$  & MeV$^{-5}$  & $1.6\cdot 10^{-2}$ & 
$4.7\cdot 10^{0}$  \\
$\gamma_{\rm S}$    & $-2.91148\cdot 10^{-17}$ & MeV$^{-8}$ & $5.2\cdot 10^{-2}$  & 
$1.3\cdot 10^{1}$  \\
$\delta_{\rm S}$    & $-4.158\cdot 10^{-10}$ & MeV$^{-4}$ & $2.6\cdot 10^{-1}$ & 
$1.8\cdot 10^{1}$ \\
$\alpha_{\rm V}$    & $2.593511\cdot 10^{-4}$ & MeV$^{-2}$ & $2.9\cdot 10^{-3}$ & 
$1.3\cdot 10^{0}$  \\
$\gamma_{\rm V}$    & $-3.8234\cdot 10^{-18}$ & MeV$^{-8}$  & $3.4\cdot 10^{-1}$  & 
$5.2\cdot 10^{1}$ \\
$\delta_{\rm V}$    & $-1.218\cdot 10^{-10}$ & MeV$^{-4}$ & $1.7\cdot 10^{0}$ &  
$6.9\cdot 10^{1}$\\
$\alpha_{\rm TS}$    & $2.34\cdot 10^{-6}$ &  MeV$^{-2}$ & $1.6\cdot 10^{1}$ & 
$2.4\cdot 10^{3}$  \\
$\alpha_{\rm TV}$    & $3.241\cdot 10^{-5}$ & MeV$^{-2}$ & $1.1\cdot 10^{0}$  & 
$1.6\cdot 10^{2}$  \\
$\delta_{\rm TV}$    & $-6.0\cdot 10^{-11}$ & MeV$^{-4}$ & $2.5\cdot 10^{1}$ & 
$3.7\cdot 10^{2}$ \\
\hline\hline
$V_{\rm P}$ & $-321$ & MeV~fm$^3$ & $8.7\cdot 10^{-1}$ & $1.6\cdot 10^0$\\
$V_{\rm N}$ & $-308$ & MeV~fm$^3$ & $8.1\cdot 10^{-1}$ &  $1.3\cdot 10^0$\\
\end{tabular}
\caption{
The set PC--F2 of coupling constants emerging from the fitting procedure
including the linear isovector-scalar term. In columns 4 and 5 the
uncorrelated and correlated errors are shown as originating from the
fitting procedure. The total $\chi^2$, $\chi^2$ per point and $\chi^2$ per 
degree of freedom are $\chi^2_{\rm tot} = 98.5$, $\chi^2_{\rm pt} = 2.10$, 
$\chi^2_{\rm df} = 2.80$.
}\label{alphats}
\end{table}

\begin{table}[htb]
\begin{tabular}{|c|c|c|c|c|} 
coupling constant &  value & dimension & uncorr. error [\%]& corr. 
error [\%]\\
\hline
$\alpha_{\rm S}$    & $-3.835796\cdot 10^{-4}$  & MeV$^{-2}$ & $2.5\cdot 10^{-3}$  & 
$9.9\cdot 10^{-1}$ \\
$\beta_{\rm S}$     & $7.6853\cdot 10^{-11}$ & MeV$^{-5}$  & $2.0\cdot 10^{-2}$ & 
$5.3\cdot 10^{0}$ \\
$\gamma_{\rm S}$    & $-2.9062\cdot 10^{-17}$ & MeV$^{-8}$ & $6.9\cdot 10^{-2}$ & 
$1.7\cdot 10^{1}$ \\
$\delta_{\rm S}$    & $-4.1797\cdot 10^{-10}$ & MeV$^{-4}$ & $2.1\cdot 10^{-1}$ & 
$2.4\cdot 10^{1}$\\
$\alpha_{\rm V}$    & $2.593357\cdot 10^{-4}$  & MeV$^{-2}$ & $3.5\cdot 10^{-3}$  & 
$1.7\cdot 10^{0}$ \\
$\gamma_{\rm V}$    & $-3.8731\cdot 10^{-18}$ & MeV$^{-8}$  & $4.4\cdot 10^{-1}$ & 
$5.9\cdot 10^{1}$\\
$\delta_{\rm V}$    & $-1.1997\cdot 10^{-10}$  & MeV$^{-4}$ & $6.9\cdot 10^{-1}$ & 
$8.0\cdot 10^{1}$ \\
$\alpha_{\rm TV}$    & $3.549\cdot 10^{-5}$ & MeV$^{-2}$ & $1.2\cdot 10^{0}$ & 
$7.6\cdot 10^{0}$\\
$\gamma_{\rm TV}$   & $-5.4\cdot 10^{-17}$   & MeV$^{-8}$ & $5.7\cdot 10^{1}$ &  
$1.8\cdot 10^{2}$ \\
$\delta_{\rm TV}$    & $-4.0\cdot 10^{-11}$  & MeV$^{-4}$ & $1.1\cdot 10^{2}$ & 
$4.0\cdot 10^{2}$\\
\hline\hline
$V_{\rm P}$ & $-321$ & MeV~fm$^3$ & $1.4\cdot 10^0$ &  $1.7\cdot 10^0$ \\
$V_{\rm N}$ & $-308$ & MeV~fm$^3$ & $1.2\cdot 10^0$ &  $1.3 \cdot 10^0$ \\
\end{tabular}
\caption{
The set PC--F3 of coupling constants emerging from the fitting procedure
including the nonlinear term in the isovector-vector density. In
columns 4 and 5 the uncorrelated and correlated errors are shown as
originating from the fitting procedure. The total $\chi^2$, $\chi^2$ per point and $\chi^2$ per degree of freedom are $\chi^2_{\rm tot} = 98.8$, $\chi^2_{\rm pt} = 2.10$, $\chi^2_{\rm df} = 2.82$.
}\label{gammatv}
\end{table}

\begin{table}[htb]
\begin{tabular}{|c|c|c|c|c|} 
coupling constant &  value & dimension & uncorr. error [\%]& corr. 
error [\%]\\
\hline

$\alpha_{\rm S}$    & $-3.83564\cdot 10^{-4}$  & MeV$^{-2}$ & $2.9\cdot 10^{-3}$   & 
$1.0\cdot 10^{0}$ \\
$\beta_{\rm S}$     & $7.6806\cdot 10^{-11}$ & MeV$^{-5}$  & $2.0\cdot 10^{-2}$ &  
$5.7\cdot 10^{0}$\\
$\gamma_{\rm S}$    & $-2.9105\cdot 10^{-17}$ & MeV$^{-8}$ & $6.9\cdot 10^{-2}$ & 
$1.9\cdot 10^{1}$ \\
$\delta_{\rm S}$    & $-4.16057\cdot 10^{-10}$ & MeV$^{-4}$ & $2.1\cdot 10^{-1}$ & 
$2.3\cdot 10^{1}$\\
$\alpha_{\rm V}$    & $2.593614\cdot 10^{-4}$ & MeV$^{-2}$ &  $3.5\cdot 10^{-3}$ & 
$1.5\cdot 10^{0}$ \\
$\gamma_{\rm V}$    & $-3.844\cdot 10^{-18}$ & MeV$^{-8}$  & $4.4\cdot 10^{-1}$  & 
$6.8\cdot 10^{1}$ \\
$\delta_{\rm V}$    & $-1.2154\cdot 10^{-10}$ & MeV$^{-4}$ & $6.8\cdot 10^{-1}$ & 
$7.6\cdot 10^{1}$ \\
$\alpha_{\rm TS}$    & $-5.92\cdot 10^{-6}$ & MeV$^{-2}$ & $7.8\cdot 10^{0}$ & 
$6.8\cdot 10^{3}$ \\
$\delta_{\rm TS}$   & $-1.12\cdot 10^{-10}$ & MeV$^{-4}$ & $4.1\cdot 10^{1}$ & 
$8.4\cdot 10^{2}$ \\
$\alpha_{\rm TV}$    & $3.937\cdot 10^{-5}$ & MeV$^{-2}$ & $1.0\cdot 10^{0}$  & 
$8.9\cdot 10^{1}$ \\
$\delta_{\rm TV}$    & $3.0\cdot 10^{-12}$ & MeV$^{-4}$ & $1.5\cdot 10^{3}$  & 
$3.1\cdot 10^{4}$\\
\hline\hline
$V_{\rm P}$ & $-321$ & MeV~fm$^3$ & $1.4\cdot 10^0$ & $1.5\cdot 10^0$  \\
$V_{\rm N}$ & $-308$ & MeV~fm$^3$ & $1.2\cdot 10^0$ & $1.4\cdot10^0$  \\
\end{tabular}
\caption{
The set PC-F4 of eleven coupling constants emerging from the fitting procedure
including 4 isovector coupling constants. In
columns 4 and 5 the uncorrelated and correlated errors are shown as
originating from the fitting procedure. The total $\chi^2$, $\chi^2$ per point 
and $\chi^2$ per degree of freedom are $\chi^2_{\rm tot} = 98.2$, 
$\chi^2_{\rm pt} = 2.09$, $\chi^2_{\rm df} = 2.89$.
}\label{p-f4}
\end{table}

\begin{table}[htb]
\begin{tabular}{|c|c|c|c|c|c|}
coupling constant & value from NL--Z2 & PC--F1 & PC--F2 & PC--F3 & PC--F4 \\
\hline
$\alpha_{\rm S}$ [MeV]$^{-2}$& $ -4.225 \cdot 10^{-4}$  & $-3.836\cdot 10^{-4}$  & 
$-3.836\cdot 10^{-4}$        &   $-3.836\cdot 10^{-4}$      &   
$-3.836\cdot 10^{-4}$    \\
$\delta_{\rm S}$ [MeV]$^{-4}$& $-1.737 \cdot 10^{-9}$  &   $-4.185\cdot10^{-10}$  &  
$-4.158\cdot 10^{-10}$      &    $-4.180\cdot 10^{-10}$     &   
$-4.161\cdot 10^{-10}$      \\
\hline
$\alpha_{\rm V}$ [MeV]$^{-2}$&  $2.739 \cdot 10^{-4}$  & $2.593\cdot 10^{-4}$ & 
$2.594\cdot 10^{-4}$       &    $2.593\cdot 10^{-4}$    &  
$2.594\cdot 10^{-4}$      \\
$\delta_{\rm V}$ [MeV]$^{-4 }$ &  $4.502 \cdot 10^{-10}$  & $-1.192\cdot 10^{-10}$   & 
$-1.218\cdot 10^{-10}$        &   $-1.120\cdot 10^{-10}$      &   
$-1.215\cdot 10^{-10}$     \\
\hline
$\alpha_{\rm TV}$ [MeV]$^{-2}$& $ 3.566\cdot 10^{-5}$ &  $3.468\cdot 10^{-5}$ &  
$3.241\cdot 10^{-5}$       &    $3.549\cdot 10^{-5}$    &   
$3.937\cdot 10^{-5}$      \\
$\delta_{\rm TV}$ [MeV]$^{-4}$&  $6.125 \cdot 10^{ -11}$ &  $-4.20 \cdot 10^{-11}$  &  
$-6.0\cdot 10^{-11}$      &    $-4.0\cdot 10^{-11}$    &  
$3.0\cdot 10^{-12}$      \\
\hline
$\alpha_{\rm TS}$ [MeV]$^{-2}$&  --         &    --    & $2.34\cdot 10^{-6}$           
&   --     &   $-5.92\cdot 10^{-6}$      \\
$\delta_{\rm TS}$ [MeV]$^{-4}$&  --         &    --     &    --    &   --     &  
$-1.12\cdot 10^{-10}$    \\
\end{tabular}\caption{
Coupling constants from the RMF-FR interaction NL-Z2
and corresponding values from the RMF-PC interactions
PC-F1 to PC-F4.
}\label{reltowalecka}
\end{table} 

\begin{table}[htb]
\begin{tabular}{|l|r|r|r|r|r|r|}
& PC--F1 & PC--LA & NL-Z2 & NL3 & SLy6 & SkI3 \\ \hline
$\rho_0$ (fm$^{-3}$) & $0.151$ & 0.148 & 0.151 & 0.148 & 0.159 & 0.158\\
$E/A$ (MeV)  &  $-16.17$  & -16.126 & -16.07 & -16.24 & -15.90 & -15.96\\
$m^{*}/m$ & 0.61 & 0.575 & 0.583 & 0.595 & 0.690 & 0.577 \\
$K$ (MeV)  & 270  & 264  & 172 & 272 & 230 & 258\\
$a_{\rm sym}$ (MeV) & 37.8  & 37.194 & 39.0 & 37.4 & 32.0 & 34.8\\
\end{tabular}
\caption{Bulk properties of nuclear matter for the forces under consideration}
\label{nucmat}
\end{table}

\begin{table}[htb]
\setdec 0.000
\centering
%\caption{QCD-Scaled Coupling Constants for Four Relativistic Point Coupling
%Interactions}
\vspace{18pt}
\Large
\begin{tabular}{|c|c|c|c|c|l|}
Coupling Constant & $c_{ln}(\rm{PC-F1})$ & $c_{ln}(\rm{PC-F2})$ & 
$c_{ln}(\rm{PC-F3})$ & $c_{ln}(\rm{PC-F4})$ &
 Order in $\Lambda$ \\ \hline
$\alpha_{S}$ &\dec -1.641 &\dec -1.641 &\dec -1.641 &\dec -1.641 & $\Lambda^{0}$ \\
$\beta_{S}$ &\dec 1.443 &\dec 1.443 &\dec 1.443 &\dec 1.442 & $\Lambda^{-1}$ \\
$\gamma_{S}$ &\dec -2.695 &\dec -2.701 &\dec -2.696 &\dec -2.700 & $\Lambda^{-2}$ \\
$\delta_{S}$ &\dec -1.061 &\dec -1.054 &\dec -1.060 &\dec -1.055 & $\Lambda^{-2}$ \\
$\alpha_{V}$ &\dec 1.109 &\dec 1.109 &\dec 1.109 &\dec 1.109 & $\Lambda^{0}$ \\
$\gamma_{V}$ &\dec -0.360 &\dec -0.355 &\dec -0.359 &\dec -0.357 & $\Lambda^{-2}$ \\
$\delta_{V}$ &\dec -0.302 &\dec -0.309 &\dec -0.304 &\dec -0.308 & $\Lambda^{-2}$ \\
$\alpha_{TS}$ & -- &\dec 0.040 & -- &\dec -0.101 & $\Lambda^{0}$ \\
$\delta_{TS}$ & -- & -- & -- &\dec -1.134 & $\Lambda^{-2}$ \\
$\alpha_{TV}$ &\dec 0.593 &\dec 0.555 &\dec 0.607 &\dec 0.674 & $\Lambda^{0}$ \\
$\gamma_{TV}$ & -- & -- &\dec -80.470 & -- & $\Lambda^{-2}$ \\
$\delta_{TV}$ &\dec -0.422 &\dec -0.612 &\dec -0.404 &\dec 0.026 & $\Lambda^{-2}$ \\
 & & & & &  \\
\# $c_{ln}$ & 9 & 10 & 10 & 11 &  \\
\# natural & 9 & 9 & 9 & 9 &  \\
$|\rm{max}|/|\rm{min}|$ & 8.92 & 67.5 & 264.7 & 103.8 &  \\
 & & & & &  \\
$\chi^{2}_{df}$ & 2.75 & 2.80 & 2.82 & 2.89 & \\
\end{tabular}
\caption{QCD-Scaled Coupling Constants for Four Relativistic Point Coupling
Interactions.}
\label{V.T1}
\end{table}

\end{document}